\documentclass[acmsmall,nonacm]{acmart}
\usepackage{wrapfig}
\usepackage[final]{pdfpages}
\usepackage{listings}
\usepackage{xspace}

\AtBeginDocument{%
  \providecommand\BibTeX{{%
    \normalfont B\kern-0.5em{\scshape i\kern-0.25em b}\kern-0.8em\TeX}}}

\setcopyright{none}
\copyrightyear{2022}
\acmYear{2022}
\acmDOI{XXXXXXX.XXXXXXX}

\acmJournal{PACMPL}
\acmVolume{1}
\acmNumber{OOPSLA} 
\acmArticle{1}
\acmYear{2022}
\acmMonth{4}
\acmDOI{} 
\startPage{1}

\begin{document}

\title{A fast in-place interpreter for WebAssembly}

\author{Ben L. Titzer}
\email{btitzer@andrew.cmu.edu}
\affiliation{%
  \institution{Carnegie Mellon University}
  \streetaddress{4665 Forbes Avenue}
  \city{Pittsburgh}
  \state{Pennsylvania}
  \country{USA}
  \postcode{15213}
}

\newcommand\WK[1]{\texttt{\textbf{#1}}}
\newcommand\wasm[1]{\texttt{\textbf{#1}}\xspace}
\newcommand\code[1]{\texttt{\textbf{#1}}\xspace}
\newcommand\arch[1]{\texttt{\textbf{#1}}\xspace}
\newcommand\engine[1]{\texttt{\textbf{#1}}\xspace}
\newcommand\ourengine{\texttt{\textbf{wizard}}\xspace}
\newcommand\X{$\times$\xspace}
\newcommand\bench[1]{\texttt{#1}}


\begin{abstract}
  WebAssembly (Wasm) is a compact, well-specified bytecode format that offers a portable compilation target with near-native execution speed.
  The bytecode format was specifically designed to be fast to parse, validate, and compile, positioning itself as a portable alternative to native code.
  It was pointedly \emph{not} designed to be interpreted directly.
  Instead, design considerations at the time focused on competing with native code, utilizing optimizing compilers as the primary execution tier.
  Yet, in JIT scenarios, compilation time and memory consumption critically impact application startup, leading many Wasm engines to later deploy baseline (single-pass) compilers.
  Though faster, baseline compilers still take time and waste code space for infrequently executed code.
  A typical interpreter being infeasible, some engines resort to compiling Wasm not to machine code, but to a more compact, but easy to interpret format.
  This still takes time and wastes memory.
  Instead, we introduce in this article a fast in-place interpreter for WebAssembly, where no rewrite and no separate format is necessary.
  Our evaluation shows that in-place interpretation of Wasm code is space-efficient and fast, achieving performance on-par with interpreting a custom-designed internal format.
  This fills a hole in the execution tier space for Wasm, allowing for even faster startup and lower memory footprint than previous engine configurations.
  
\end{abstract}



\maketitle

\section{Introduction}

Emerging first for the Web in 2017~\cite{WasmPldi}, WebAssembly is a portable, low-level compilation target supported in all major browsers.
Originally designed as a successor to asm.js~\cite{AsmJs}, which allowed C/C++ to be compiled to JavaScript, it supplanted other technologies such as Native Client~\cite{NativeClient} as the new best target for native compilation to the Web.
Since that time, WebAssembly has seen rapid uptake in a number of new spaces, including cloud computing~\cite{FemtoCloud}, digital contracts, edge computing~\cite{WasmEdgeDancer}\cite{FastlyEdge}, IOT~\cite{WasmIotOs}, and embedded systems~\cite{WarDuino}.

A key design criteria for Wasm was offering performance competitive with native code.
Initially, top-tier performance was considered paramount, and approaching native code performance to compete with technologies like Native Client was realized by reusing the optimizing JIT compiler infrastructure in browsers.
Yet Wasm bytecode was also designed to be fast to parse, verify, and compile--criteria validated during the design process by building single-pass validators and single-pass decoding to SSA compiler IR to minimize upfront costs in the compilation pipeline.

However, despite minimizing bytecode parsing work by careful design, optimizing compilers inescapably take considerable time and memory to produce good native code, penalizing application startup in JIT scenarios.
To address startup time problems, browsers prototyped separate, faster compilers during Wasm's design phase, validating that the same choices that enabled single-pass verification enabled single-pass compilation.
Such often-termed ``baseline'' compilers spend far less compilation time, often 10\X - 20\X less, but produce code that typically runs 1.5\X to 3\X slower than an optimizing compiler.
This represents a classic tradeoff space known to VMs for decades; more compilation time means better code quality.
Today, all browser engines employ multiple Wasm compiler tiers to strive for \emph{both} good startup time \emph{and} high throughput.

\subsection{Whither the Interpreter?}

Seemingly overlooked in this development arc is the obvious choice of using an \emph{interpreter} to execute bytecode.
After all, traditionally, virtual machines are developed with an interpreter first.
There are a lot of advantages to interpreters.

\begin{enumerate}
  \item Since interpreters are easier to write, understand, and maintain, they allow more rapid experimentation in bytecode design.
  \item Since they need no translation or rewriting step, start up is fast.
  \item Bytecode is usually more compact than machine code, so interpeters generally use less memory than compilers.
  \item Debugging application code is easier, as the interpreter loop can be stopped at any instruction and program state inspected, altered, and resumed.
  \item Interpreters are easier to audit, since there is a fixed amount of code, and usually have fewer security vulnerabilities.
  \item Dynamic code generation is sometimes impossible, either because is not allowed on the platform, like iOS, or code space is limited. 
\end{enumerate}

For all these reasons, nearly all virtual machines, from pioneering work on Lisp to Smalltalk to Self, to today's broadly-accepted VMs such as the Java Virtual Machine, the CLR, Python, Ruby, and JavaScript, have an interpreter.

Why then, is Wasm any different?
The answer is simply that efficient interpretation was explicitly \emph{not} in the design criteria\footnote{In fact, in a smoky back room, I probably declared, ``Interpreters don't matter here.''}.
But some Wasm engines do indeed employ interpreters, such as JavaScriptCore, Chakra\footnote{Though discontinued, ChakraCore was the first Wasm engine to feature a rewriting interpreter.}, and Wasm3.
These engines use interpreters for exactly the advantages listed above.
Yet none of these interpreters work directly on the original bytecode; all of them rewrite Wasm bytecode to a different internal format.
Rewriting Wasm bytecode has similar disadvantages to baseline-compiling: it still takes time and memory.

\subsection{The Final Tier is Shed}

There is an important point missing in the Wasm virtual machine design tradeoff space.
An \emph{in-place} interpreter, i.e. one that interprets the original bytes of a binary module, would offer the best startup time and lowest memory consumption.
For cold or never executed code, where the downside of the interpreter's much slower execution speed is outweighed by the major savings of avoiding translation time, it would be the optimal choice.
Employed in concert with compilation tiers for hot code, such an interpreter would serve the role it does in other mature systems like JVMs.
Is it possible to interpret Wasm in-place efficiently?
Until now, this was thought infeasible.
In this work, we solve this open problem and supply the missing point in the design tradeoff space: the first fast in-place interpreter for Wasm (empirical measurements in Figure~\ref{fig:execution_xlate_time}\&\ref{fig:execution_xlate_space}).

\noindent
\begin{minipage}[t]{.5\textwidth}

  \begin{figure}[H]
    \centering
    \caption{Execution time vs translation time.}
  \includegraphics[width=2.6in]{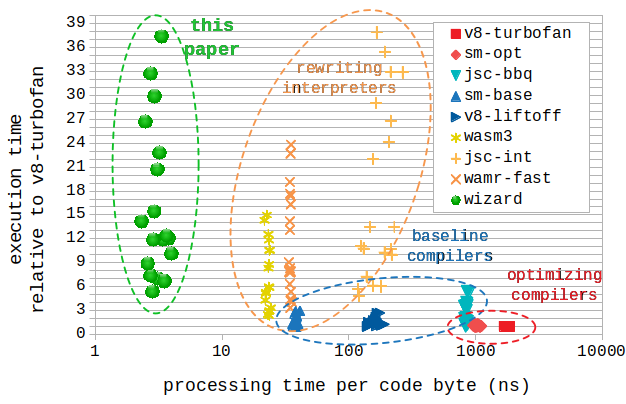}
    \label{fig:execution_xlate_time}
\end{figure}

\end{minipage}
\begin{minipage}[t]{.5\textwidth}
  
  \begin{figure}[H]
    \centering
    \caption{Execution time vs translation space.}
  \includegraphics[width=2.6in]{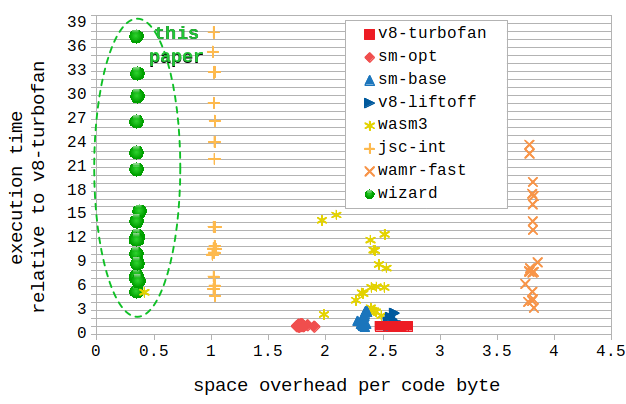}
    \label{fig:execution_xlate_space}
\end{figure}
  
\end{minipage}

\vspace{12pt}

We first identify the interpreter-crucial information that is missing in Wasm's bytecode design.
This information, namely key control-flow and value stack information, is not actually missing, but rather \emph{implicit}.
Our insight is that the validation algorithm for Wasm bytecode \emph{already} computes this information in its modeling the control and value stack during typechecking.
All that remains is to distill a few key numbers into a compact side-table that is used during interpretation.
The side-table is organized so that all accesses occur in constant ($O(1)$) time, and no searches of the table are necessary.
Thus the interpreter always has relevant information directly at hand and behaves like a standard interpreter.
Other details are important for making a fast interpreter as well, such as hand-coding key parts in assembly and combining exactly the right layout of value stack and virtual memory protections to robustly handle application stack overflow without needing any explicit checks.

With these new techniques, we have finally achieved an in-place interpreter for Wasm that is on par with state-of-the-art interpreters for other bytecodes and for Wasm interpreters using rewriting.
This paper completes the triad of basic tier designs (interpreter, fast compiler, optimizing compiler) for Wasm.
In a comical twist of fate, Wasm's tiers have arrived exactly backward!


\subsection{Organization}

The remainder of this paper is organized as follows.
Section~\ref{sec:wasm} recaps the design of Wasm's functions, stack machine, and control flow constructs, which are key to understanding why interpretation has been challenging until now.
Section~\ref{sec:sidetable} shows the design of the side table used for the interpreter and how the validation algorithm already contains the key information necessary to emit the side table in a single pass.
Section~\ref{sec:interpreter} details the interpreter implementation, including key assembly techniques to achieve the best-performing dispatch loop, and the design of the data structures necessary to make a fast operand and execution stack.
Section~\ref{sec:evaluation} evaluates the interpreter on standard benchmarks and compares translation time, memory consumption, and execution time to JITs (\ref{sec:tiers}) and other interpreters (\ref{sec:interpreters}) for Wasm.
Prior work related to optimizing interpreters is summarized in Section~\ref{sec:related}, followed by the conclusion.

\section{Wasm Design}\label{sec:wasm}

Wasm provides a low-level programming model consistent with its original goal of a minimal, high-performance abstraction over hardware.
Its principal features include:
\begin{itemize}
\item \wasm{i32}, \wasm{i64}, \wasm{f32}, \wasm{f64}, and \wasm{v128} primitive types
\item an opaque reference type
\item 32- and 64-bit integer arithmetic
\item single- and double-precision floating point arithmetic
\item 128-bit vector operations
\item large byte-addressable memories with explicit load and store instructions
\item functions with local variables
\item direct and indirect function calls
\item global variables
\end{itemize}

The Wasm binary format is designed to be compact yet fast to decode and validate in a single pass.
This includes not just bytecode, but all constructs.

\subsection{Modules and Instances}

Wasm code is organized into \emph{modules} which are in turn organized into a list of \emph{sections}.
Sections in a module declare functions, memories, tables, global variables and static data.
Bytecode is grouped into \emph{functions} with statically-typed \emph{parameters}, \emph{results}, and \emph{local variables}.
All operations in core Wasm manipulate only a module's own internal state.
Modules must \emph{import} functions (and memories, tables, etc) in order to access platform capabilities or state outside the module.
Imports may be provided by the ``host'' environment, such as JavaScript and the Web, or from other modules.

A module is akin to an executable file, or part of one, rather than an executing program.
To run, a module must be \emph{instantiated}, supplying bindings for its imports.
At instantiation time, a Wasm engine creates the state (tables, globals, and memories) declared by the module, with the result being called an \emph{instance}.
An instance may \emph{export} its own functions, memories, tables, etc. to other modules or the host environment.

The primary dynamic storage of a Wasm program is typically one large, bounds-checked, byte-addressable \emph{memory}, but global variables and tables of opaque host references can also be used.
Future proposals will add first-class function references and garbage-collected objects to Wasm.
These too are forms of local state and must be shared explicitly with other instances.

\subsection{Bytecode design}

\textbf{Functions.}
All code\footnote{other than trivial initializers for globals} in Wasm is organized into functions.
Functions each have a signature with a fixed number of parameter and result types, such as \wasm{[i32 f32 externref] -> [i32 i32]}.
Execution of a Wasm program entails executing functions that may call each other, maintaining an execution stack\footnote{Note that the execution stack is not aliased by Wasm memory, thus not vulnerable to stack smashing} that stores their local variables and operands, and running their internal code.
In a binary module, the body of a Wasm function begins by declaring the number and type of their additional local variables, followed immediately by the bytecode.

\textbf{Stack machine.}
As is common for many bytecode designs, Wasm is a \emph{stack machine}, meaning individual bytecode instructions take their operand values from an \emph{operand stack} and push their results back.
Local variables are separate.
To be used, a local must be explicitly loaded onto or stored back from the operand stack.
Implementations typically store them internally as a prefix of the operand stack, together referred to here and throughout as the \emph{value stack}.
The arguments of an outgoing function call become the first locals of the callee function.

\textbf{Structured control flow.}
Unlike most bytecode designs, however, Wasm has \emph{structured control flow constructs} such as blocks, ifs, loops, and switches that are encoded inline in the bytecode.
We refer to them as \emph{structured}, since they must be properly nested.
This was a deliberate choice for compactness and to ensure that bytecode validation can be done in a single pass with minimal data structures\footnote{Provably minimal, if a CFG is restructured from the known algorithm for optimal interval analysis, ensuring the validation metadata per control flow construct is discarded and reused as promptly as possible.}.
In contrast, a typical bytecode design with jumps usually requires more bytes to store and two passes to verify.

\textbf{Direct interpretation not straightforward.}
Most bytecode formats can be interpreted directly in their binary form (i.e. in-place), with an instruction pointer stepping through the bytes of the original code.
Jumps typically have an offset or instruction number of the target instruction directly in the bytes, allowing a constant-time adjustment of the instruction pointer.
But Wasm is unusual in that a branch instruction specifies a target construct by \emph{relative nesting depth}, transferring control to the beginning (in case of \wasm{loop}), middle (in the case of \wasm{if}) or end (for \wasm{block}, \wasm{if}, \wasm{else}) of the construct.
Wasm is also unusual in that branches can also copy and pop values off the operand stack\footnote{Other stack machine designs, like JVM bytecode, allow values on the operand stack while branching, but stack heights and contents must match. Thus JVM branches cannot implicitly pop values.}.
Thus a direct Wasm interpreter faces two unusual problems when executing a branch:
\begin{itemize}
\item can't quickly find the target bytecode offset, e.g. start of a \wasm{loop} or \wasm{end} of a \wasm{block}
\item can't determine how many values to pop off the operand stack
\end{itemize}

To understand how this paper efficiently solves this open problem, we must first journey deep into how Wasm bytecode validation is done in production-quality engines.

\subsection{Bytecode validation}

Wasm, though low-level, is typed.
A number of static checks ensure that a module is well-formed.
Within a function, all instructions, including arithmetic, calls, control flow, etc. must be applied to the correct number and type of operands.
All control flow constructs must be properly nested with no invalid branches.
In the specification, this validation is expressed in a standard type system formalism.
In engines, the algorithm is implemented as an abstract interpreter that models an \emph{abstract} control and value stack.
We describe such an implementation here to later make it clear how our modifications affect an already very efficient verification implementation.

\begin{figure}
  \includegraphics[width=\textwidth]{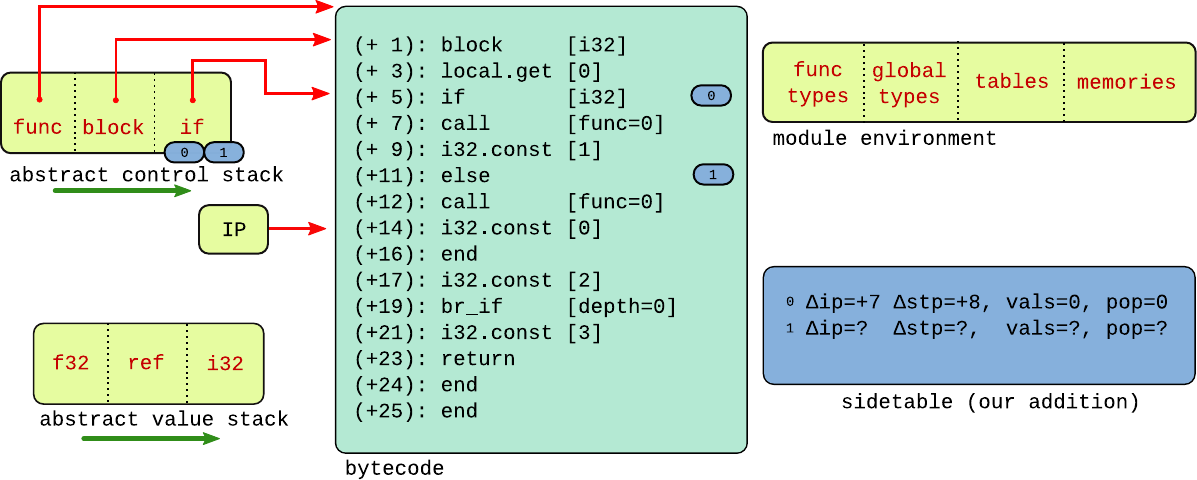}
    \caption{Illustration of data structures used in production Wasm code validators.}
    \label{fig:code_validation}
\end{figure}

Figure \ref{fig:code_validation} illustrates the operation of a production quality Wasm code validator.
Three primary data structures are used.
\begin{itemize}
\item The \emph{module environment} models the types of functions, tables, globals, and number of memories of the enclosing module.
  The module environment is not mutated during verification of a function's bytecode.
  \item An \emph{abstract control stack} models the nested control-flow constructs, keep tracking of where each starts and its expected parameter and result types\footnote{In Wasm, all control constructs can have data parameters and results, represented as a block signature. Thus a block can be seen as a super-instruction; it pops values off the stack, and every path to its end pushes the same number and type of results. This allows for very compact code, often obviating local variables. In practice, most blocks have an empty block signature.}.
    This is used to properly match \wasm{block}, \wasm{if}, and \wasm{loop} starts with their \wasm{else} and/or \wasm{end}.
  \item An \emph{abstract value stack} tracks abstract values for stack slots and local variables.
    In current Wasm, abstract values are simply \emph{value types}, i.e. \wasm{i32}, \wasm{i64}, \wasm{externref}, etc.
    If a future proposal introduces flow-sensitive validation, the abstract values for locals would need to be extended to include initialization information.
\end{itemize}
    
The validation algorithm proceeds in a single forward pass over the bytecode, never needing to backtrack.
For simple instructions like arithmetic or calls which only pop their operands from the value stack and push results back, the algorithm pops and checks required input types and pushes resulting output types.
Control-flow instructions are validated using the control stack.
For example, a \wasm{br} (branch) instruction references the target \wasm{block} or \wasm{loop} by relative nesting depth; the validator matches the opening construct by indexing into the control stack.
Since, in Wasm \wasm{block}, \wasm{if}, and \wasm{loop} can have parameter and result types, the validator must check that the value stack contains the appropriate types expected for the target construct.

Importantly, an \wasm{end} bytecode closes a control-flow construct, either a \wasm{block}, \wasm{if}, \wasm{loop}, or the whole function.
At \wasm{end}, all branches that can legally target the construct have been seen.
Thus the implicit target bytecode offset of all branches \emph{to this construct} are known, because that offset must be either the start of a \wasm{loop} or, for any other construct, the \wasm{end}, i.e. \emph{here}.
This information just needs to be saved somewhere easily accessible for the interpreter.
After processing \wasm{end}, the algorithm pops the control stack entry and can reuse its storage space, which is optimally efficient.

\section{Interpreter Design}\label{sec:interpreter}

In this section, we present our fast in-place Wasm interpreter design.
\newline
The key enabling techniques are:
\begin{itemize}
  \item an innovative side-table design which allows efficient access to missing branch information
  \item a highly efficient value stack organization for $O(1)$ local variable and operand access as well as zero-copy function calls
\end{itemize}

Additionally, we chose to implement the core logic of the interpreter in hand-written assembly language\footnote{More precisely, a macro assembler API in a high-level language that has methods to generate individual instructions, allowing it to be configurable in ways that typical textual assembly languages are not.}, which allows for near-perfect register allocation and unlocks all possible dispatch and organization techniques.
We discuss the rationale for hand-written assembly at the end of this section.

\subsection{Sidetable Design}\label{sec:sidetable}

As we've seen, Wasm control-flow bytecodes represent nested control constructs, rather than low-level jumps.
However, an interpreter needs the bytecode offset of where to go if a branch is taken, ideally in $O(1)$ time.
This includes not only explicit branches like \wasm{br}, \wasm{br\_if}, \wasm{br\_table}, but also the implicit branch in \wasm{else}.
Note that \wasm{block}, \wasm{loop}, \wasm{end} and \wasm{return} are \emph{not} branching bytecodes, since they either fall through or exit the function.

To supply the in-place interpreter with the missing information, the validation algorithm saves a portion of its work into a per-function side-table data structure separate from the original bytecode.
This side-table is a compact, highly-efficient mapping from branch origin to target offset, plus some additional stack manipulation information.

\noindent
\begin{minipage}[t]{.5\textwidth}

  \begin{figure}[H]
  \includegraphics[width=2.2in]{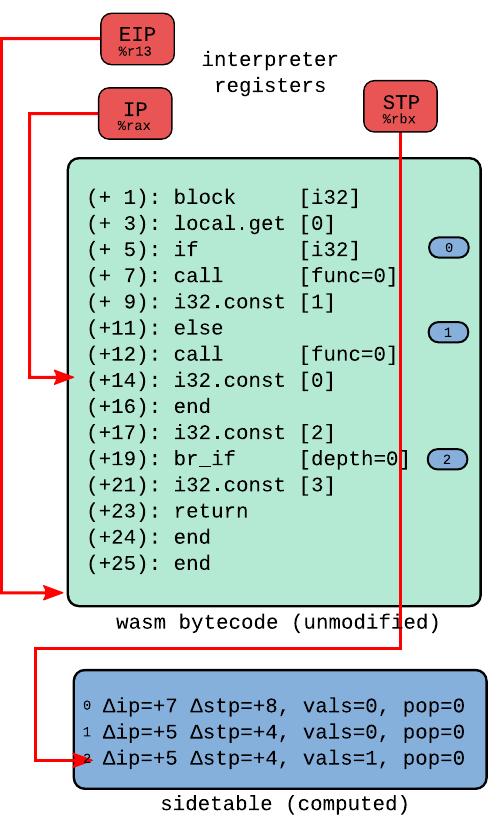}
    \caption{Interpreter code and sidetable.}
    \label{fig:sidetable}
\end{figure}

\end{minipage}
\begin{minipage}[t]{.5\textwidth}

  \lstdefinestyle{interp}{
    basicstyle=\linespread{0.67}\ttfamily\footnotesize
}

\lstset{style=interp}

\begin{lstlisting}[caption={Interpreter pseudocode for branches.},label={lst:sidetable},captionpos=b]
  function_entry:
    IP  := &func->code->original[0];
    STP := &func->code->sidetable[0];
    decodeLocals();
    executeNextBytecode();
  
  handle_BR:
    doControlTransferFromSTP();
    
  handle_ELSE:
    doControlTransferFromSTP();
  
  handle_BR_IF:
    var cond = popI32();
    if (cond != 0) {
      doControlTransferFromSTP();
    } else {
      IP = nextIP;
      STP = STP + #STP_entry_size;
      executeNextBytecode();
    }
  
  handle_BR_TABLE:
    var key = popI32();
    var max = STP->maxcase;
    if (key > max) key = max;
    STP = STP + (1 + key) * #STP_entry_size;
    doControlTransferFromSTP();

  def doControlTransferFromSTP() {
    moveValues(STP->valcnt, STP->popcnt);
    IP = IP + STP->delta_ip;
    STP = STP + STP->delta_stp;
    executeNextBytecode();
  }
    
\end{lstlisting}

\end{minipage}

\vspace{8pt}

To make the data structure time- and space-efficient, it consists of entries sorted by branch origin and omits non-branch instructions.
It is emitted as a side-effect of the single-pass validation algorithm above.
Because the validation algorithm already visits bytecodes in forward order, it simply emits branch entries as it goes, obviating a separate sorting step.
Since only branches need entries, the sidetable is very small, often empty.
Empirically, most Wasm functions are small with no control flow, so they have no side-table at all.

Every entry in the side-table is a 4-tuple of the form $\langle$$\Delta$\code{ip}, $\Delta$\code{stp}, \code{valcnt}, \code{popcnt}$\rangle$, where:

\begin{itemize}
\item $\Delta$\code{ip} the amount to adjust the instruction pointer by if the branch is taken
\item $\Delta$\code{stp} the amount to adjust the \emph{side-table pointer} by if the branch is taken
\item \code{valcnt} the number of values that will be copied if the branch is taken
\item \code{popcnt} the number of values that will be popped if the branch is taken
\end{itemize}

\subsubsection{The Sidetable Pointer}

Like most interpreters, our in-place interpreter maintains an instruction pointer (\code{IP}) into the bytecode during execution.
It also maintains an \emph{end} instruction pointer (\code{EIP}), which is used to check if the program falls off the end of the function, which is a legal implicit return in Wasm.
To use the side-table, the interpreter simply maintains another pointer, the side-table pointer (\code{STP}), consulted when executing branches.

Figure \ref{fig:sidetable} illustrates the interpreter state for the bytecode and sidetable, and Listing~\ref{lst:sidetable} illustrates how side-table entries are used by the interpreter during execution.
The instruction pointer (\code{IP}) is initialized to point directly into the bytecode and the sidetable pointer (\code{STP}) points at the first side-table entry.
Branch instructions make use of the \code{doControlTransferFromSTP()} subroutine which adjusts both the \code{IP} and \code{STP} based on the entry to which it points, as well as adjusting the value stack.
Notice that a conditional branch that is not taken still must update the \code{STP} so that it points at the next entry for the next branch, if any, in the code.
Note that \wasm{br\_table} works much like a jump table, computing an index into the side-table and using the corresponding entry.

\subsection{Value stack design}

Since Wasm bytecode constitutes a stack machine, nearly all instructions access the value stack, making it crucial for interpretation speed.
A single indirection to local variables and the top of the operand stack is ideal.
Wasm's numbering scheme for locals is inspired by JVM bytecode.
Parameter \#0 is local \#0, followed by declared locals, then the operand stack.
Outgoing arguments of a call become the first local variables of the callee function's value stack.
The JVM chose this design so interpreters could \emph{overlap} the value stack of the callee frame with the operand stack of the caller, avoiding copying any argument values.
This never panned out for Wasm until now.
Our design succeeds in using the JVM trick to avoid copying arguments, but requires separating the value stack from the execution stack.

\begin{figure}
    \centering
  \includegraphics[width=\textwidth]{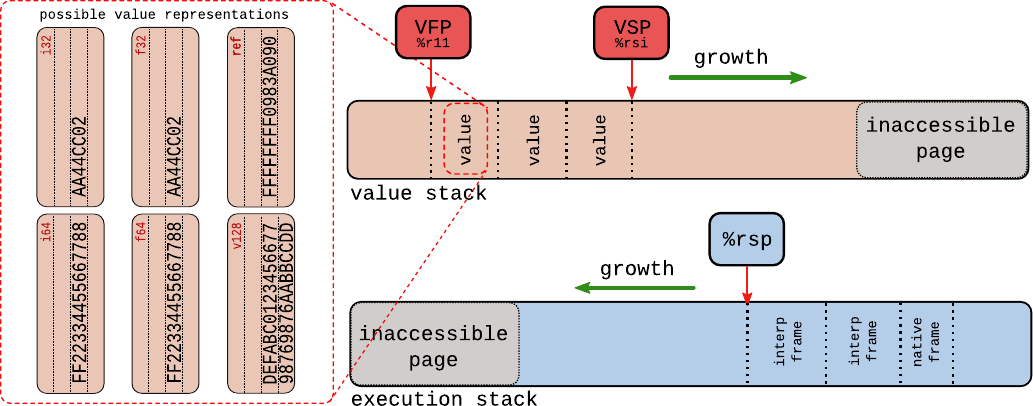}
    \caption{Value stack and execution stack layout.}
    \label{fig:value_stack}
\end{figure}

Figure~\ref{fig:value_stack} illustrates the value stack design.
We separate the storage of Wasm program values from the control information of the interpreter itself.
That is, the value stack contains only Wasm values, while the interpreter's call stack contains only control information, organized into one execution frame per Wasm call frame.
As such, the value stack is a contiguous array of Wasm values which increases in size towards higher addresses.
Though invisible to Wasm programs, and orthogonal to our design here, the interpreter frames are on the native stack and use the native stack pointer (e.g. \code{\%rsp} on \arch{x86-64}) which grows towards lower addresses.

\subsubsection{Value tagging}

Wasm values can be 32-bit or 64-bit integers, 32-bit or 64-bit floats, 128-bit SIMD vectors, or external references.
We choose to store all Wasm values in the value stack as \emph{unboxed}, so that the interpreter never needs to implicitly allocate a heap object\footnote{Boxing is a major overhead in dynamic language implementations and would be prohibitively expensive for Wasm, outweighing all optimizations described in this paper.}.
This obviously makes sense for all primitive values, since they can share storage as raw bytes in the memory of the value stack.

However, an engine may need to find \wasm{externref} values in a value stack during garbage collection.
Natively-compiling Wasm engines use stackmaps~\cite{GcStackmaps}.
But an interpreter is different, and typically cannot afford either space or time to precompute stackmaps.
Instead, we chose to tag value stack entries with a 1-byte \emph{type tag}, inflating entries to be 32 bytes, i.e. \emph{really fat}\footnote{Aligning value stack entries on a power-of-two boundary allows for shift-based arithmetic when indexing.}.
Type tags are written into the value stack only when necessary (i.e. when initializing locals in the function prologue, not at all for primitive arithmetic, etc).
To be clear, type tags are never needed for any dynamic check, since Wasm code is statically typechecked.
They can be omitted if there is no garbage collector or it uses conservative stack scanning~\cite{ConservativeGc}.

\subsubsection{Stack overflow}

Wasm engines must be robust to call stack overflow.
The Wasm specification includes one test with unbounded recursive function calls.
Every engine must fail gracefully, though the exact point where overflow is detected is not specified.

Checking for stack overflow should not ruin performance.
A check on every value stack push would be prohibitive.
Instead, we use a \emph{guard page} at the end of the value stack and rely on an OS-level signal upon fault to catch and report stack overflow.
A single guard page suffices, as the interpreter cannot inadvertantly stride over it; by design it never accesses arbitrarily far ahead.
Similarly, interpreter execution frames (on the native stack) are all fixed size and another guard page\footnote{Using \code{sigaltstack} on POSIX platforms for signal handling.} will cause a fault if the native execution stack overflows before the value stack.

\subsubsection{Putting it all together}

\begin{figure}
    \centering
  \includegraphics[width=\textwidth]{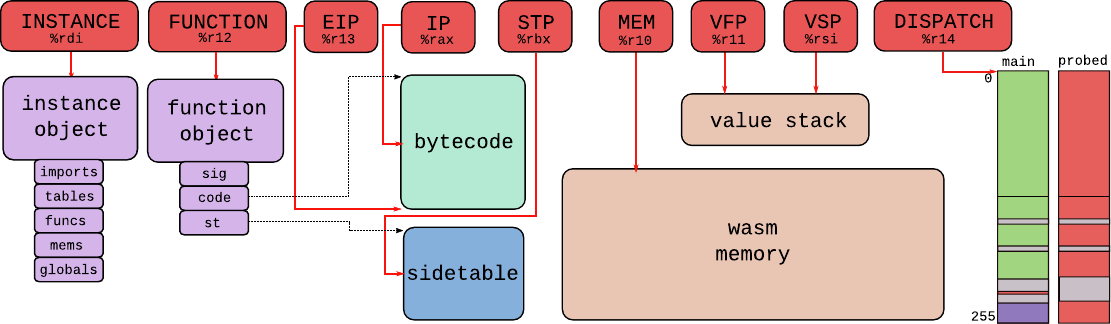}
    \caption{All interpreter registers.}
    \label{fig:interpreter_state}
\end{figure}

At this point, we've designed two critical data structures necessary to make an efficient in-place interpreter.
Figure~\ref{fig:interpreter_state} completes the set of 9 state registers used by our interpreter implementation.
In addition to the 3 ``control'' registers pointing into the bytecode and sidetable and the two ``data'' pointers into the value stack, the interpreter also requires:

\begin{itemize}
\item \code{MEM} a pointer to the start of the wasm memory\footnote{Though not shown, Wasm memories also have a guard region, obviating the need for an explicit bounds check.}
\item \code{FUNCTION} a reference to the current function
\item \code{INSTANCE} a reference to the current \emph{instance}
\item \code{DISPATCH} for dynamically enabling per-instruction probing
\end{itemize}

\subsection{Interpreter implementation in assembly}

We chose to implement our interpreter in a new Wasm research engine, \ourengine~\cite{WizardEngine}.
While most of the engine is written in a portable, safe, high-level language~\cite{VirgilPldi}, we use custom hand-written assembly for the fast interpreter.
Using assembly or a custom code generation facility is relatively common for production interpreters.
For example, the Java HotSpot virtual machine generates its interpreter~\cite{HotspotInt} using a macro assembler (at startup), V8's Ignition~\cite{Ignition} interpreter is generated from hand-written TurboFan compiler IR (at build time), and JSC's LLint is written in a macro assembler language (build time).
Several factors impinge on our decision.
\begin{enumerate}
\item an interpreter is a small, important piece of code
\item the bytecode format and semantics of Wasm are very stable
\item compilers have trouble generating optimal code for interpreter loops
\item we wish to study interpreter performance in detail
\item key dispatch techniques are difficult to obtain from compilers
\item developing and debugging assembly code is relatively time consuming
\end{enumerate}

Key advantages of using assembly to implement an interpreter are 1) its code is not perturbed by changing compiler optimizations, 2) key interpreter state can be fixed to specific architectural registers, 3) threaded~\cite{ThreadedCode}, indirect-threaded~\cite{IndirectThreadedCode}, and other dispatch techniques can be used, 4) handlers can be ordered and aligned cache line boundaries, 5) fast- and slow-paths can be organized inline or out-of-line, 6) all hardware instructions can be used, 7) self-modifying code and dispatch table swap techniques can be used, 8) error handling and hard cases can be factored from handlers, 9) very small resultant code footprint.

The interpreter that we present in this paper makes use of nearly all of these techniques.
It is implemented using a macro assembler that generates \arch{x86-64} machine code, has many switches to enable different features, and provides an instrumentation interface for profiling and debugging.
It consists of 2800 source lines of code which generate approximately 14KiB of machine code and 7KiB of dispatch tables.

\subsubsection{Dispatch tables and handlers}

\begin{figure}
    \centering
  \includegraphics[width=\textwidth]{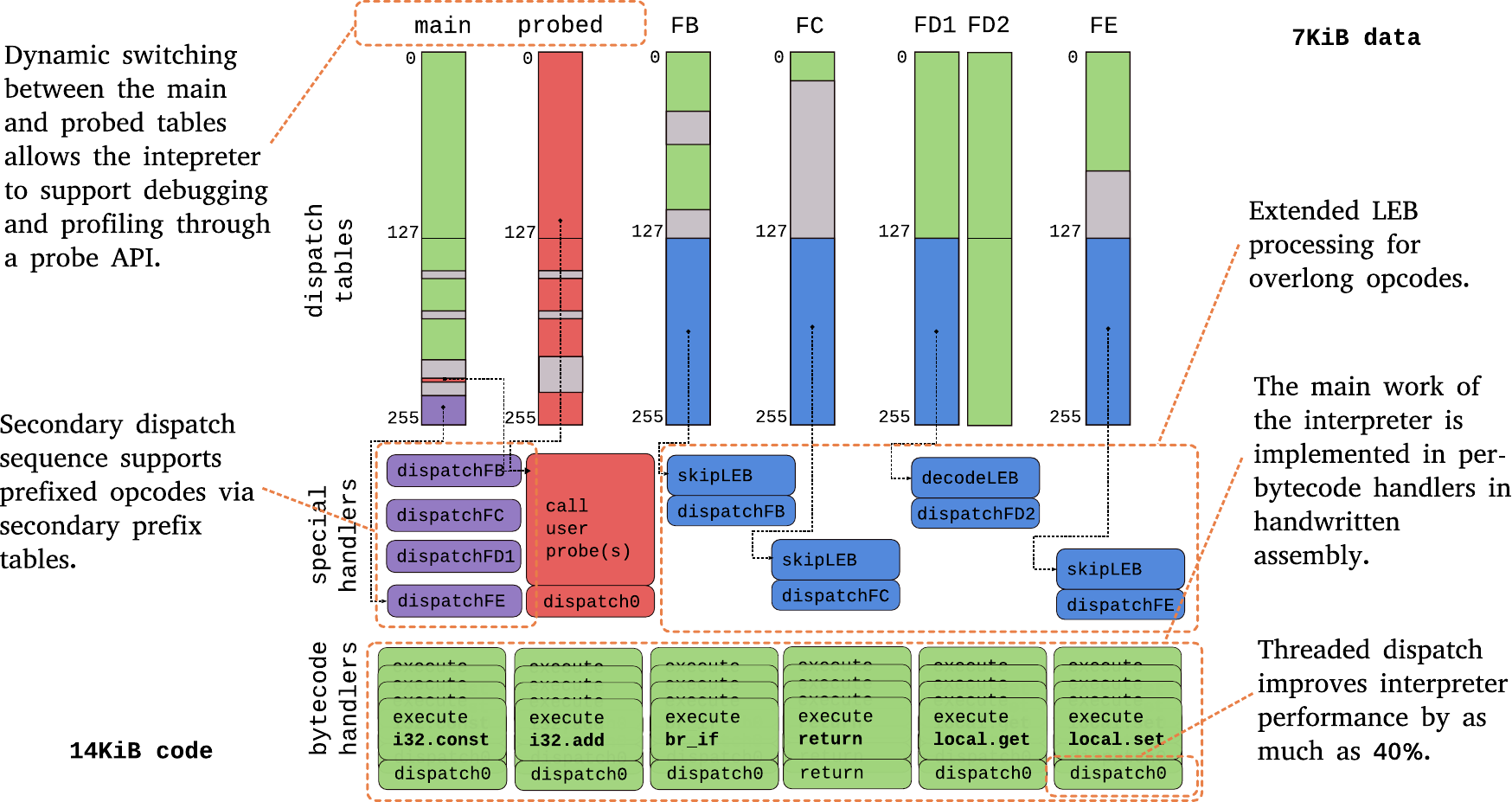}
    \caption{Dispatch tables and handlers in the fast interpreter.}
    \label{fig:dispatch_tables}
\end{figure}

Wasm is a bytecode in the true sense of the word.
An instruction is encoded as a byte-sized opcode, followed by zero or more immediates.
Longer instructions are preceded by prefix byte.
It is natural therefore to design a software interpreter around 256-entry \emph{dispatch tables} that contain pointers to \emph{handlers} that implement each bytecode.
Figure~\ref{fig:dispatch_tables} illustrates the dispatch table and handler organization for our fast Wasm interpreter.

The fast interpreter uses multiple dispatch tables, each of which points to a sequence of machine code called a \emph{handler}.
A \emph{dispatch} through a table consists of loading the address at a particular index and indirectly jumping to that address.
The first dispatch table, the \emph{main} dispatch table, is used for the first byte of an instruction.
Since the most important bytecodes in Wasm were assigned a non-prefixed opcode, the first dispatch through the main table normally lands in a handler that will directly execute the bytecode.


Prefix dispatch tables handle the tricky but rare corner cases.
If the first byte of an instruction is a prefix, then the target address in the main table will be a \emph{special handler} that loads the next byte in the instruction stream and dispatches through the appropriate table.
There is still one more wrinkle, however.
Wasm is unusual in that the opcode after a prefix can be a variable-length LEB~\cite{LEB128}, where the uppermost bit of the byte indicates continuation bytes follow.
Thus, in prefix dispatch tables, entries for the upper 128 opcodes (i.e. where the upper bit is 1) point to another special handler that fully decodes the LEB and finally dispatches to an actual bytecode handler, using yet another dispatch table.
Of current Wasm bytecodes, only the SIMD opcodes occupy the upper part of their prefix space (\code{0xFD}).
Thus, in practice, Wasm opcodes normally require just one dispatch, two if prefixed, and maximum three if the LEB opcode is longer than 1 byte.

\subsubsection{Direct vs threaded dispatch}

Each bytecode handler consists of machine instructions that manipulate the value stack or module instance or perform control flow.
A handler usually simply leaves the interpreter registers ready to start the next bytecode pointed to by \code{IP}, except for instructions like \wasm{unreachable} and \wasm{return}, which will terminate execution or return from the current interpreter frame, respectively.

Conceptually an interpreter has a single loop that repeatedly dispatches one instruction at a time, each handler jumping back to the loop header.
Nowadays, a technique known as \emph{threaded dispatch}~\cite{VmShowdown} is often used, where instead of a jump, a copy of the dispatch code is inlined at the end of each handler.
This is difficult to do in a high-level language\footnote{
Compilers generally will not transform a straight-forward loop-over-switch into threaded dispatch, as the necessary transformation, tail-duplicating the loop header for many hundreds of cases, is highly specific.
Instead, we must arduously fill out manual dispatch tables and either use tail calls or, in C, the non-standard \code{gcc} ``labels as values''~\cite{GccLabels} extension.
The resulting control flow is complex, with multiple irreducible loops and hundreds of indirect branch targets, and may lead to spills across important instructions.
To use tail calls, we must rewrite the entire interpreter as individual handler functions that end with a tail-call to the next handler (as is done in Ignition~\cite{Ignition}), passing the entire interpreter state forward as arguments.
Yet only some C compilers support a non-standard tail-call optimization.
In any case, without a custom calling convention, we run out of registers and spill some interpreter state on the stack unnecessarily.
}, but easy in assembly.
This saves a jump instruction and typically makes better use of the CPU's indirect branch predictor.
We implemented both and report the performance difference in Section~\ref{sec:evaluation}.

\subsubsection{Debugging and instrumenting with probes}

One of the motivating advantages for an interpreter tier in any virtual machine is ease of debugging and instrumentation.
In particular, an interpreter implements an exact bytecode-by-bytecode emulation of the machine, offering the possibility of stopping before or after any bytecode and inspecting the virtual machine state.
This is key to support both high and low-level debugging of a program as well as instrumentation such as profiling.

Our fast interpreter design has provisions for general instrumentation at the bytecode level.
It offers the ability to insert \emph{probes}~\cite{NonIntrusive} at either \emph{local} locations in a program, or \emph{global}, the interpreter loop itself.
Both have different use cases.
A probe is simply a callback to engine-level code that \emph{fires} when the given bytecode is executed, or each time the interpreter loop is executed.
In a probe callback, one can inspect virtual machine state through an engine API and then indicate if the program should resume normally or do something else (typically, terminate).
Probes are primitives from which debugging support (e.g. breakpoints), tracing support (e.g. logging), or profiling support (e.g. counters), are built.

Figure~\ref{fig:dispatch_tables} shows how the fast interpreter supports probes.
\begin{itemize}
\item \textbf{Local probes}.
  For a probe inserted \emph{at a particular instruction}, the original bytecode of the function containing the location is copied and the bytecode is overwritten with a special, normally-illegal bytecode, \code{PROBE}.
  Since illegal bytecodes will be rejected by the code validator, they will never appear in valid programs.
  Dispatching on \code{PROBE} will land in a special handler that looks up the user's callback associated with this bytecode, calls it, and then after, loads the \emph{original} bytecode and dispatches through the \code{main} table.
\item \textbf{Global probes}.
  For a probe inserted \emph{into the global interpreter loop}, the interpreter switches \emph{modes}, using the \code{probe} dispatch table for every instruction.
  Similar to the \emph{local} probes, the interpreter looks up the global instrumentation, calls it, and then after, dispatches through the \code{main} table.
\end{itemize}
  
It's important to note that this design allows probes to be inserted and removed dynamically.
This allows maximum flexibility to instrumentation code while allowing the interpreter to run at full speed otherwise.
For example, suppose a user wants to trace the execution of just one particular function in a module.
They could insert global instrumentation into the main interpreter loop and filter out all callbacks where the function of interest is not on the top of the stack.
But this is inefficient; the interpreter will go through the \code{probed} table every time, issue calls into the runtime, into the user code, which inspects state, etc.
A more efficient way is to insert a \emph{local} probe into the interesting function.
When the local probe is fired, it dynamically inserts the global probe, thereby getting called for every subsequent bytecode.
When the function calls another, or when it returns, the global probe can be disabled, and everything goes back to full speed.
The interpreter is careful to switch back to the \code{main} dispatch table whenever it detects global instrumentation is disabled, so it will always run fast when it has opportunity to do so.

\subsection{Tuning}

Considerable work has gone into designing efficient data structures and dispatch tables for the fast interpreter presented here.
We'd be remiss in not enumerating the many other tuning strategies applied here and what we've learned.
This was made easier by the fact that the fast interpreter presented here was implemented not in textual assembly language, but in a macro assembler framework we built in a high-level language.
Thus configuration options were easy to introduce into the code that generates the interpreter, rather than relying on macro facilities in a textual assembly language.

\begin{itemize}
\item \textbf{Manual register allocation}.
  The fast interpreter state consists of 9 registers (Figure~\ref{fig:interpreter_state}).
  All of today's 64 bit architectures have enough architectural registers that it is simply a matter of assigning interpreter registers to architectural registers.
  For \arch{x86-64}, we chose register assignments carefully to eliminate \code{REX} prefixes for the most commonly-occurring registers.
  We did not measure alternative assignments, but suspect that CPU register renaming makes additional tuning moot.
\item \textbf{Minimal dispatch sequence}.
  In addition to the choice between threaded and non-threaded dispatch, we experimented with dispatch table designs where entries were 2, 4, and 8 bytes.
  In the 2 byte design, entries are not direct addresses, but deltas that are applied to the start of a code region, requiring an additional \code{add} instruction in the dispatch sequence.
  In the 4 and 8 byte alternatives, the entries are direct addresses, constrained to be in the lower 4gb of address space in the former case.
  Of these alternatives, the \textbf{4 byte sequence is fastest}, often as much as 10\% faster.
  Unsurprisingly, we found \textbf{threaded dispatch is fastest}, on average 14\%, maximum 29\% faster.
\item \textbf{Value tagging}.
  Our interpreter design uses value tags to find GC roots when necessary.
  We evaluated the alternative and found that \textbf{eliminating tags improves performance}, on average 8\%, maximum 15\% faster.
\item \textbf{Really fat values}.
  SIMD values are 128 bits wide.
  When coupled with value tagging, values occupying 32 bytes of memory each in the value stack.
\item \textbf{Inline/out-of-line LEB decoding}.
  The interpreter must dynamically decode the many immediates in Wasm that are encoded as variable-length LEBs, e.g. the index of a \wasm{local.get}.
  Often these variable-length immediates are just a single byte.
  We experimented with moving the uncommon case out-of-line and concluded that \textbf{out-of-line LEB slow cases} could be as much as 5\% faster.
\item \textbf{Memory \#0 base pointer}.
  Wasm programs access memory very frequently.
  As described earlier, our implementation caches a pointer to the base of the (first) Wasm memory in an architectural register.
  We did not evaluate the performance impact of this optimization. 
\item \textbf{Handler alignment}.
  Bytecode handlers are short but critically important sequences of code.
  We suspected they may be subject to microarchitectural effects of instruction and trace caches arising from code alignment. 
  We experimented with 1, 2, 4, 8, 16, and 32 byte alignment of handlers but \textbf{detected no statistically significant performance variation.}
\item \textbf{Handler order}.
  We suspect that handler code order may introduce significant microarchitectural effects~\cite{WrongData}~\cite{Stabilizer}.
  However, we did not study the effect of handler order beyond simply emitting common bytecode handlers first. 
\item \textbf{Error case sharing}.
  Several Wasm bytecodes \emph{trap} on error cases, like divide by zero, unrepresentable floats, etc.
  We factored the error handling paths in order to save space.
  Since traps usually terminate the program, the performance does not matter.
\item \textbf{Call/entry/exit sharing}.
  We exploit commonalities among call bytecodes (\wasm{call}, \wasm{call\_indirect}, and tail calls\footnote{From the \wasm{tail-call} Wasm proposal.}), \wasm{return}/fall-through, as well as branches in order to save code space.
  This may have a small effect on performance, but we did not measure this.
\item \textbf{Handler sharing}.
  We exploit the fact that some instructions end up with identical handler code and share the handlers (e.g. \code{block} and \code{loop}, some memory stores).
  We did not measure alternatives.
\end{itemize}

The above conclusions are supported by a performance evaluation of alternatives that is beyond the scope of this paper.
To summarize those experiments, the overall difference between the best (tuned) and worst (untuned) interpreter performance is 20\% to 60\% across the benchmark suite.
Interestingly, as we will see in the next section, this difference is enough to make our interpreter design meet and exceed existing (re-writing) interpreters in comparative performance tests.

\section{Evaluation}\label{sec:evaluation}

In this section, we evaluate our Wasm interpreter against many state-of-the-art Wasm engines.
The goal is to assess our claim that an in-place interpreter supplies the missing point in the execution tier tradeoff space between translation time, space overhead, and execution time.
In particular, an in-place interpreter should offer superior translation time and space overhead compared to other tiers.
Ideally, such an interpreter should also have comparable execution time to existing interpreter tiers.
Of course we expect that interpreters \emph{should be} handily outclassed by JIT compilers for long-running programs, so we shouldn't expect to \emph{replace} them.
Our experiments quantify the tradeoff space empirically.

\subsubsection{Wasm Engines in 2022}\label{sec:engines}

Today, five years after support was announced in 4 major browsers, engines are significantly different, and many new competitors have appeared.
In particular, Web engines have evolved significantly from what has been reported in the literature to date.
Today, all Web engines are sophisticated multi-tier systems.
\begin{itemize}
\item \engine{V8}~\cite{V8} -
  two compiler tiers.
  V8 eagerly compiles the entirety of a module with Liftoff~\cite{Liftoff}, a baseline compiler, with many parallel threads. Upon completing baseline compilation, a module is ready to run.
  Optimized compilation with TurboFan~\cite{TurboFan}, the optimizing compiler shared with JavaScript, is begun in parallel in the background.
  Optimized code gradually replaces baseline code as it is finished, until all functions in a module are fully optimized.
  Either compiler can be disabled with command-line flags.
\item \engine{SpiderMonkey}~\cite{Spidermonkey} -
  two compiler tiers.
  Similar to V8, with a baseline compiler first compiling a whole module and then an optimizing compiler in the background, patching in optimized code as it is completed.
  Either compiler can be disabled with flags.
\item \engine{JavaScriptCore} (\engine{JSC})~\cite{JavaScriptCore} -
  one interpreter and two compiler tiers.
  Wasm modules are not eagerly compiled.
  Instead, individual functions are lazily translated to interpreter bytecode for LLint~\cite{LLint}.
  Dynamic counters tier-up hot functions from interpretation to \engine{BBQ}, a fast compiler that has a minimal IR, and then to \engine{OMG}, a fully optimizing compiler based on B3~\cite{B3}.
  Either compiler can be disabled with flags.
\item \engine{ChakraCore}~\cite{ChakraCore} -
  one interpreter tier and one compiler tier.
  Now largely unsupported, Wasm modules were not eagerly compiled.
  Instead, individual functions were lazily translated to interpreter bytecode.
  Dynamic counters tier-up hot functions from interpretation to a fully optimizing compiler.
\end{itemize}

In addition to the rapid evolution of Web engines, a number of non-Web Wasm engines have appeared.
\begin{itemize}
\item \engine{wasmtime}~\cite{Wasmtime} - A standalone runtime implemented in Rust.
  \newline
  One compiler tier.
  The optimizing Cranelift compiler can be used in JIT or AOT modes.
\item \engine{WAVM}~\cite{WAVM} - A standalone runtime implemented using LLVM.
  \newline
  One compiler tier.
  Consists primarly of a Wasm loader and the LLVM compiler backend and can only be used in AOT mode.
\item \engine{wamr}~\cite{Wamr} - (WebAssembly Micro Runtime) a lightweight standalone runtime.
  \newline
  Two interpreter tiers and one compiler tier, but only one at a time.
  The ``classic'' interpreter, we discovered, is an in-place interpreter that does not use a side table, but a control flow cache.
  The ``fast'' interpreter is a rewriting interpreter that reuses most of the standard Wasm bytecodes but rewrites control flow.
  Both interpreter loops are written in C and use \code{gcc} extensions for threaded dispatch, if possible.
  A one-pass JIT can be used instead.
\item \engine{wasm3}~\cite{Wasm3} - The self-proclaimed fastest WebAssembly interpreter\footnote{It is; we confirm their measurements in our experiments. Hey, but speed isn't everything.}.
  \newline
  One interpreter tier.
  All design considerations emphasize interpreter performance.
  It is implemented using tail calls and relies on \code{gcc} tail-call optimization.
  The internal bytecode format consists of a list of function pointers to handlers, plus immediates, i.e. classic threaded code.
\item \engine{wasmer}~\cite{Wasmer} - A standalone runtime for lightweight containers.
  \newline
  Three compiler tiers, one at a time.
  Packages two compilers from other projects: Cranelift (from \engine{wasmtime}) or LLVM, and offers its own one-pass compiler.
\end{itemize}

\subsubsection{Our Tier Choices}

We chose only to compare against execution tiers that perform \emph{dynamic} translation, intentionally omitting those engines performing static (AOT) translation.
Further, though this paper focuses on interpreter design, we include several experiments that compare engines with multiple tiers to understand their current tradeoffs.
We therefore measure these execution tiering configuration:

\begin{itemize}
\item \ourengine - Our engine with its in-place interpreter that uses a sidetable.
\item \engine{wamr-classic} - The \engine{WAMR} engine with its ``classic'' in-place interpreter.
\item \engine{wamr-fast} - The \engine{WAMR} engine with its (now default) rewriting interpreter.
\item \engine{wasm3} - Default configuration of a rewriting interpreter.
\item \engine{v8-liftoff} - \engine{V8} with only the Liftoff baseline JIT.
\item \engine{v8-turbofan} - \engine{V8} with only the TurboFan optimizing JIT.
\item \engine{sm-base} - \engine{Spidermonkey} with only the baseline compiler.
\item \engine{sm-opt} - \engine{Spidermonkey} with only the optimizing compiler.
\item \engine{jsc-int} - \engine{JSC} with JITs disabled, i.e. only the rewriting interpreter.
\item \engine{jsc-bbq} - \engine{JSC} with the rewriting interpreter and \engine{BBQ} quick JIT only.
\item \engine{jsc-omg} - \engine{JSC} with the rewriting interpreter and \engine{OMG} optimizing JIT only.

\end{itemize}

\subsection{Benchmark setup}

All tests were performed on a Linux 4.15 kernel on a machine with 32GiB of RAM and one Intel Core i7-8700K CPU @ 3.7GHz and the CPU governor set to ``performance''.
Benchmarks used are the PolyBenchC-4.2.1 suite, with the MEDIUM dataset.
We used V8 version \texttt{10.2.0}, JSC version (roughly 2022-02-01), and Spidermonkey version \texttt{C101.0a1}, all release builds built from source.
Data was collected in two experiments; 100 uninstrumented runs of 10 engine configurations on the 24 benchmarks to gather execution time, and 100 instrumented runs of the same to collect translation time and space statistics.
Every run represents a separate OS process.
Execution times are of the complete process (i.e. not internally timed), while translation times are the sum over all Wasm functions translated in the run.
All timing results are the average over the 100 runs, and when error bars are shown, they represent the 5\textsuperscript{th} and 95\textsuperscript{th} percentiles of the distribution.

\subsection{Translation Time}

All of our chosen execution tiers, except \ourengine and \engine{wamr-classic}, the only other in-place interpreter of which we are aware, apply some form of translation to the input bytecode.
Rewriters either generate an internal bytecode (interpreters) or machine code (compilers).
Thus we measure the time taken for the respective translation by instrumenting the source code of each engine by adding time and space measurements.
In the case of \engine{wamr-fast}, the custom bytecode is generated as a side-effect of validation, so we measure the \emph{additional} time for translation by subtracting the baseline validation time obtained from \engine{wasmr-classic}, which has no translation time.
\ourengine doesn't \emph{translate}, but instead the reported time is the sidetable tracking and construction time, measured as the difference between validation time with and without sidetable tracking.

\begin{figure}
\includegraphics[width=\textwidth]{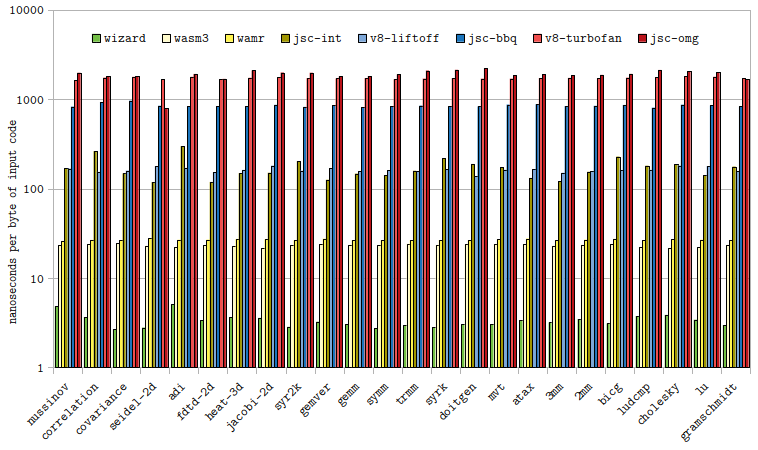}
    \caption{Translation time normalized to bytes of input code.}
    \label{fig:translation_time}
\end{figure}

We plot the average translation time, divided by the number of input bytes translated. 
The ratio of translation time to input bytes normalizes differences in tiering strategy (e.g. lazy compilation) and benchmark size.
Figure~\ref{fig:translation_time} gives the results (note Y-axis is logarithmic).

Here we focus on configurations that isolate individual tiers, rather than multi-tier adaptive configurations.
The experimental results show a dramatic difference in translation time for these tiers, nearly 3 orders of magnitude.
Though baseline compilers often differ from optimizing compilers by more than 10\X, they are still 10\X more expensive than rewriting interpreters.
Yet there is overlap, as the rewrite time of \engine{jsc-int}, an interpreter, is almost on-par with \engine{v8-liftoff}, a baseline compiler.
Similarly, the \engine{jsc-bbq} quick compiler is closer to an optimizing compiler, as it uses an IR, unlike \engine{v8-liftoff}.
The winner is clearly our in-place interpreter design, as the sidetable generation in \ourengine is a full order of magnitude cheaper than the cheapest rewriting interpreters.

\subsection{Translation Space}

\begin{figure}
\includegraphics[width=5in]{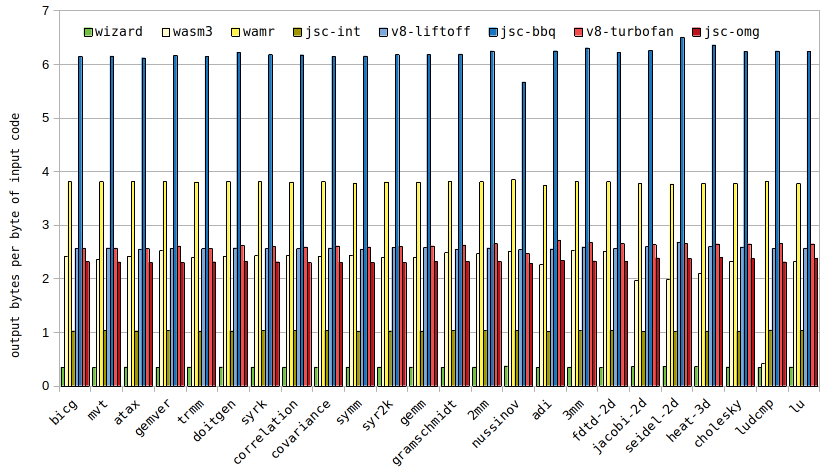}
    \caption{Translation output bytes normalized to bytes of input code.}
    \label{fig:translation_space}
\end{figure}

Translation not only consumes time, but space.
We measure the memory overhead of translation in terms of bytes generated per input byte of code.
Here again, the ratio of output bytes to input bytes normalizes differences in tiering strategy (e.g. lazily compilation) and benchmark size.
Note that we only measure the size of generated code (whether internal bytecode or machine code) and not additional metadata such as code objects or a source position table, which all rewriters need for debugging.
We also do not measure per-module space costs.
Thus this experiment \emph{underestimates} space costs of rewriters when debugging is involved.
Figure~\ref{fig:translation_space} gives the results.

There are several somewhat surprising results here.
We find that some rewriting interpreters (such as \engine{wamr}) can consume up to 4\X as much space as the original bytecode, while others (such as \engine{jsc-int}) consume about the same amount of space as the original bytecode.
We believe this is because \engine{jsc-int} uses an internal bytecode similar to JSC's JavaScript bytecode, which has been heavily tuned to reduce memory consumption on webpages.
Also somewhat surprising is that JIT compilers, which generate machine code, do not necessarily consume more code space, in general, than rewriting interpreters, though \engine{jsc-bbq} is an outlier.
We discovered that \engine{wasm3}, like \engine{wamr}, often trades space for time--nearly all of its bytecode quantities are word-sized.
Yet during rewriting, \engine{wasm3} does a number of peephole-like optimizations, globally canonicalizes constants, and uses a register machine internally, all of which save space.

These measurements show the sidetable in \ourengine takes far less space, only about 30\% additional space compared to the original bytecode,
as it only requires entries for control-flow\footnote{Note that here, we measure \ourengine sidetable entries compressed to 2 bytes where possible, though this is not the default.}, and many sidetables are empty.
That is a full order of magnitude more space-efficient than the other tiers, which typically cost 2\X to 4\X the original bytecode's space.

\subsection{Execution time}

Figure~\ref{fig:exec_time} gives the absolute execution times of the benchmarks for the \engine{v8-turbofan} and \engine{wasm3} execution tiers.
Execution times reported in other figures in this section are normalized to either of these two baselines.

\begin{figure}[b]
  \tiny
\begin{tabular}{r | r r r r | r r r r | r r}
 benchmark & \engine{v8-turbofan} & \engine{wasm3} & & benchmark & \engine{v8-turbofan} & \engine{wasm3} & & benchmark & \engine{v8-turbofan} & \engine{wasm3} \\\hline
\bench{bicg} & 0.016571 & 0.007197 & & \bench{covariance} & 0.023764 & 0.072963 & & \bench{3mm} & 0.037620 & 0.203351 \\
\bench{mvt} & 0.016617 & 0.007252 & & \bench{symm} & 0.023545 & 0.087849 & & \bench{fdtd-2d} & 0.029838 & 0.204593 \\
\bench{atax} & 0.016593 & 0.007344 & & \bench{syr2k} & 0.024594 & 0.106093 & & \bench{jacobi-2d} & 0.031257 & 0.244652 \\
\bench{gemver} & 0.016756 & 0.009553 & & \bench{gemm} & 0.024285 & 0.105961 & & \bench{seidel-2d} & 0.153029 & 0.349776 \\
\bench{trmm} & 0.020411 & 0.042696 & & \bench{gramschmidt} & 0.028422 & 0.117820 & & \bench{heat-3d} & 0.038734 & 0.363499 \\
\bench{doitgen} & 0.021046 & 0.054735 & & \bench{2mm} & 0.029670 & 0.122010 & & \bench{cholesky} & 0.078381 & 0.773620 \\
\bench{syrk} & 0.020491 & 0.069131 & & \bench{nussinov} & 0.032324 & 0.189651 & & \bench{ludcmp} & 0.088084 & 0.893945 \\
\bench{correlation} & 0.023820 & 0.068604 & & \bench{adi} & 0.071388 & 0.228267 & & \bench{lu} & 0.088218 & 0.876272 
\end{tabular}
\caption{Execution time of benchmarks (in seconds) to which Figures~\ref{fig:jit_vs_interpreter}\&\ref{fig:interpreter_time} are normalized.}
\label{fig:exec_time}
\end{figure}

\subsubsection{Interpretation vs JIT compilation}\label{sec:tiers}
\begin{figure}
\includegraphics[width=\textwidth]{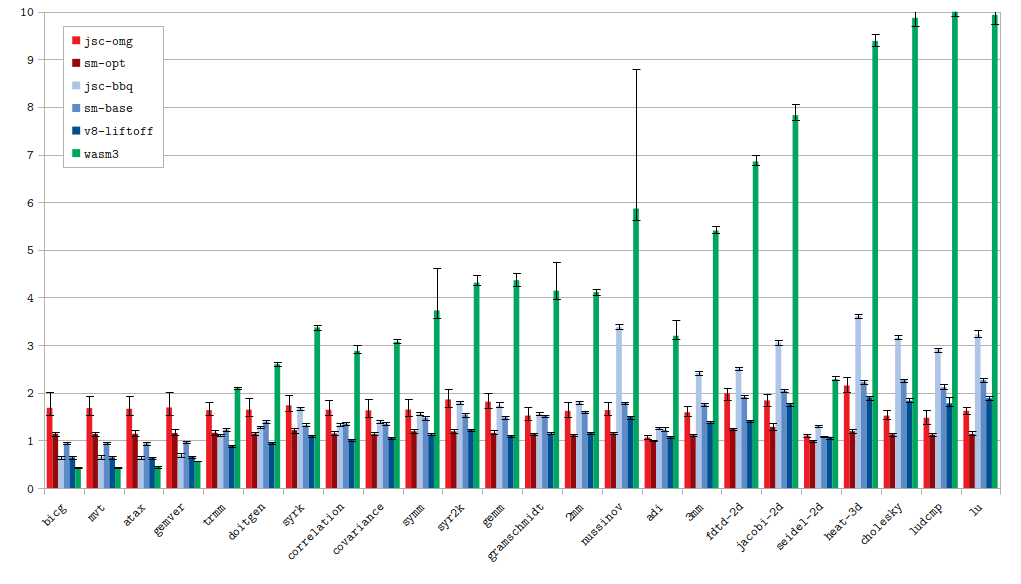}
    \caption{Execution time (relative to \engine{v8-turbofan}) of JITs versus the fastest Wasm interpreter.}
    \label{fig:jit_vs_interpreter}
\end{figure}

Interpreters start up faster, but have slower execution time compared to JIT compilers.
But how much?
We measured the execution time of the benchmarks in the six single-JIT tier configurations and the \emph{fastest} interpreter that we measured, \engine{wasm3}.
Since the execution time of benchmarks varies by two orders of magnitude, we normalize execution times for each benchmark to the corresponding execution time on \engine{v8-turbofan}.
Figure~\ref{fig:jit_vs_interpreter} gives the results.

On the horizontal axis, we sorted the benchmarks by their execution time on \engine{wasm3} (same as in the table).
As we can see, the shortest-running benchmarks on the left side of the graph do not run long enough to benefit from the work spent by the optimizing compiler, and all baseline compilers are faster, with the interpreter fastest.
There is also a fixed cost of starting a relatively heavyweight JSVM, which contributes to the interpreter being fastest for the shortest 4 benchmarks.
Moving to the right, as execution time increases with longer-running benchmarks, the fixed cost of startup and the cost of compilation are increasingly amortized.
Thus the middle of the graph shows more balanced results; the interpreter falls behind but the baseline compilers, particularly \engine{v8-liftoff}, remain competitive.
Continuing on, the gap between optimizing compilers (particularly \engine{v8-turbofan} and \engine{sm-opt}) and the rest increases; execution time is now dominated by code quality.
Baseline compilers level off, with \engine{v8-liftoff} and \engine{sm-base} around 2\X slower and \engine{jsc-bbq} closer to 3\X.
The common rule of thumb that interpreters are 10\X slower than JITs turns out to be roughly accurate in the end, as there is a clear trend towards roughly 10\X for \engine{wasm3} vs \engine{v8-turbofan} here.
These results match our intuition, but better, \emph{quantify} it, giving us fairly round numbers to reason with.
They also reaffirm the need to have a broad picture of execution time:
\begin{itemize}
\item Each of (interpreter, baseline compiler, optimizing compiler) runs some benchmark fastest.
\end{itemize}

\subsubsection{Interpreter showdown}\label{sec:interpreters}

\begin{figure}
\includegraphics[width=\textwidth]{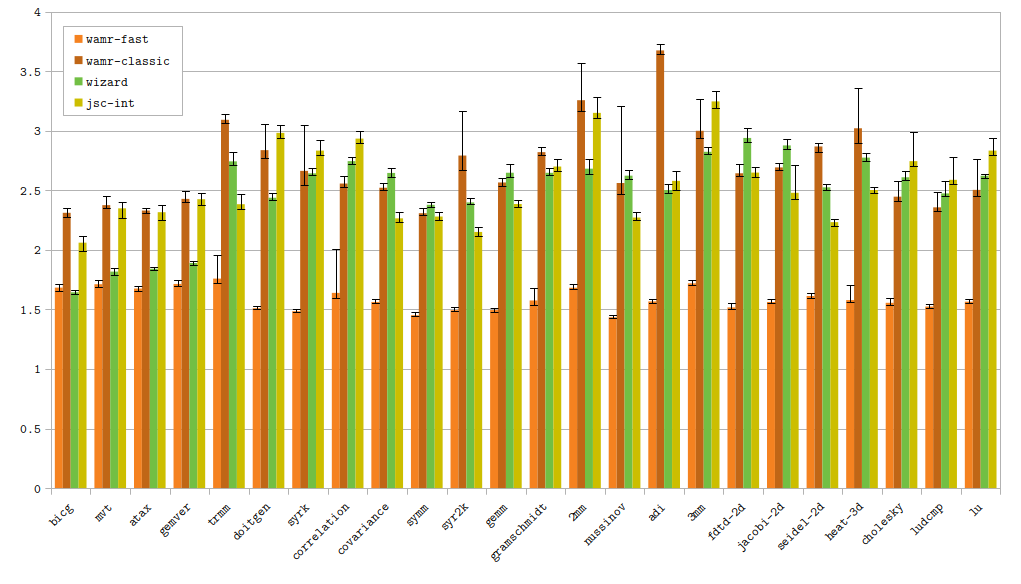}
    \caption{Execution time (normalized to \engine{wasm3}) of interpreters.}
    \label{fig:interpreter_time}
\end{figure}

Of course, this paper focuses on making a fast, in-place interpreter.
Using similar benchmarking methodology, on the same benchmarks, we evaluated the five interpreter tiers.
Figure~\ref{fig:interpreter_time} gives the results.

As we found that \engine{wasm3} was consistently the fastest interpreter across the board, we chose to normalize all execution times in this graph to it.
Here we find that \engine{wamr-fast}, the configuration of the \engine{wamr} engine with its rewriting interpreter, is consistently 2\textsuperscript{nd} fastest, nearly always 1.5\X to 1.7\X slower than \engine{wasm3}, while the others are typically 2\X - 3\X slower.
We attribute this not only to \engine{wasm3}'s threaded code dispatch technique, but its several bytecode-level optimizations.
Our design, \ourengine, performs better than the other in-place interpreter design, \engine{wasm-classic} in most situations.
It performs nearly equivalent to \engine{wamr-fast} on the 4 shortest-running benchmarks.

\subsubsection{Multi-tier configurations in context}

Note that our experiments in this section focused on isolating individual execution tiers in order to study their tradeoffs.
In production configurations, all web engines run in multi-tier configurations, as described in Section~\ref{sec:engines}.
The space for tiering designs is vast, with many variables, such as laziness, concurrency, thresholds for tiering up, on-disk caching, etc.
We ran additional experiments for engines in their default configuration, but we found the the results are complex enough to merit study in their own right--a topic that is unfortunately beyond the scope of this paper.

\subsubsection{Avoiding pathological behavior}

We discovered \engine{wamr-classic} uses an in-place design just days before publication.
Since then, we studied its implementation in earnest.
It is written in C and uses \code{gcc} extensions to construct a jump table for threaded dispatch.
The jump table is crucial for performance; disabling it via a configuration variable reduces performance by more than 2\X, which reaffirms the need for this crucial optimization in interpreter design.
Of course, we tested its fastest configuration, as 2\X would have put it handily out of the running.

However we discovered that, instead of an $O(1)$ sidetable like our design, it uses both a \emph{dynamic control stack} and a cache of control entries that help it find branch targets during runtime.
The cache is a fixed-size, 128-element, 2-way set associative cache indexed by branch address.
That amounts to a fixed cost per module, rather than per function (we did not include its space cost in Figure~\ref{fig:translation_space}).
A cache miss results in a slow path where the entire function may be rescanned to the end, repopulating the cache with new entries, including an entry for the current branch.
The performance could be pathological for a large program with many branches, a fact obscured by the relatively small benchmark programs in our suite.
To verify the pathological behavior, we constructed an adversarial program consisting of thousands of branches, approximating a larger program, and observed slowdowns of as much as 8\X.
This vulnerability was known to its authors and is one of the reason that \engine{wamr} now ships the rewriting interpreter (\engine{wamr-fast}) by default.

Given the potential pathological behavior of \engine{wamr-classic}, we still believe that \ourengine represents the first viable \emph{fast} in-place design.

\section{Related Work}\label{sec:related}

Interpreters are as old as the hills.
From the first popular interpreted language, Lisp~\cite{Lisp} in 1960, to today's modern scripting and data manipulation languages like Python, Ruby, R, PHP, and MatLab, interpreter performance has been a key subject of interest.
Many languages that are now fast through dynamic compilation were once primarily interpreted, such as Smalltalk, Self, Java, C\#, OCaml, and JavaScript.
Python is still most-often interpreted, and its performance is still of key concern~\cite{PythonInt}~\cite{QuantifyingPythonInt}.
Because of their advantages, new interpreters continue to appear, even for languages previously compiled~\cite{DatalogInterp}.

\subsubsection{Fixed or flexible format?}
Research into interpreter techniques either assumes the code format to be \emph{fixed}, such as standardized bytecode formats like the JVM, the CLR, Dalvik, and WebAssembly, where binary programs arrive metaphorically chiselled into stone, or \emph{flexible}, where the format can be changed to suit the needs of a specific language or implementation.
Clearly the larger design space of flexible formats affords more techniques, though key lessons hopefully improve the design of future fixed formats.
For example, a key question of interest is whether a stack machine (such as the JVM or WebAssembly), or a register machine (such as the CLR or Dalvik bytecode), is inherently more efficient~\cite{VmShowdown}~\cite{VirtRegMachine}.
Though it is too late to change the JVM, CLR, Dalvik, or Wasm, research here informs the next proposed format.

\subsubsection{Interpreter dispatch techniques.}

As our own experimental results reaffirm, the dispatch sequence is critical to interpreter performance.
If the format is flexible, e.g. if entirely in-memory, then threaded code~\cite{ThreadedCode} or more compact indirect threaded code~\cite{IndirectThreadedCode} can be used.
A large amount of work has aimed to improve the predictability of indirect branches~\cite{BranchPredInt} by exploiting the microarchitectural details of BTBs~\cite{BTBs}.

\subsubsection{Superinstructions.}

For flexible formats, the number of dispatches can be reduced with superinstructions~\cite{Superinstructions}, replacing small opcodes with larger, combined opcodes.
This can be done online\cite{DynamicSuperInst}, offline, or a mixture of both. 
Super-instructions can be formed from combinations of simpler instructions, often in embedded Java VMs~\cite{JavaCardCompress}, or by defining language-specific high-level operations like in CPython and JavaScript VMs.

\subsubsection{Radically different interpreter IRs.}
Bytecode is not the only interpretable format.
Abstract syntax trees or other intermediate representations can be interpreted directly in memory.
Usually, speed is not the goal, though recently a fast AST-walking interpreter has been described for R~\cite{RAstInt}.
Instead, interpreting IR has other benefits, e.g. proving the correctness of the interpreter, partial evaluation, and collapsing multiple levels of interpretation~\cite{CollapsingInt}.
Truffle/Graal~\cite{SelfOptimizingAst} uses ASTs, for example \cite{GraalSqueak}, as a way to express an interpreter for the Futamura~\cite{Futamura} projection.
Other graph-based IRs have been explored~\cite{DynamicInterp}.
Some compilers have IR interpreters~\cite{Lli} for testing.
A standard compilers course~\cite{Cs132} includes interpreters for each IR during translation.
None seem to compete with bytecode interpreters in speed.

\subsubsection{Optimizing fixed format interpreters.}
A number of interpreter optimizations can still apply to fixed formats.
The most common is duplicating the dispatch sequence at the end of every handler (referred to in this paper and elsewhere, if somewhat imprecisely, as a threaded interpreter or threaded dispatch), and is used in all the interpreters we tested.
Threaded dispatch has even been applied to hosted interpreters on the JVM~\cite{HostedInterpreters}.
For stack machines, stack caching~\cite{StackCaching}~\cite{CombiningStackCaching}~\cite{JavaIntOpt} VMs try to keep the top-of-stack cached in a register, reducing loads and stores to the value stack.
It is possible to further duplicate~\cite{CodeCopyingVM} handlers to get more BTB entries~\cite{OpIndirectAccuracy}.
Recent work has also proposed new hardware support for indirect speculation~\cite{HwSwCodesign}~\cite{JavaOnCell} or to directly address BTB entries~\cite{ShortCircuitDispatch}.
Many JVMs mutate bytecode in-place~\cite{Quickening} to replace symbolic references with indexes and offsets.

\subsubsection{Interpreter generators.}

Writing a highly efficient interpreter remains a black art.
It is challenging, sometimes impossible, to convince compilers for high-level languages to emit perfect interpreter code.
A number of interpreter generation frameworks have been proposed, including Tiger~\cite{Tiger}, VMGen~\cite{vmgen}, and a JavaScript VM generator~\cite{JsvmGen}, that automate much of the tedious bookkeeping and broadly apply best practices.

\subsubsection{Interplay with JITs and debugging.}

If the virtual machine is allowed to generate new machine code, then the entire VM design space is unlocked.
Context threaded code~\cite{MixedModeContextThreading}, where code is represented as a sequence of calls to handlers, can be used.
From there, selective inlining~\cite{OptDirectThreadInline} 
of handlers can be applied to improve performance.
Discussion of JIT designs is beyond the scope of the paper, yet their interactions with interpreters is of note.
Trace compilation~\cite{TraceCompilation} is fed by collecting execution traces from an interpreter.
Meta-tracing~\cite{MetaTracing}, i.e. tracing through the handler implementation, has been employed by the PyPy dynamic optimizer.
Dynamic adaptive optimization with deoptimization is now standard in many virtual machines. 
The design of interpreter stack frames determines the cost and complexity of on-stack replacement. 
In-place interpretation has been shown to ease debugging support, since no mapping need be maintained from rewriting. 

\section{Conclusion}\label{sec:conclusion}

In-place interpretation has long been considered when designing fixed code formats such as the JVM, CLR, and Dalvik, since it is the most memory-efficient.
Wasm is unique in that it was explicitly designed with near-native performance as the highest priority and engines shipped with optimizing compiler tiers first.
Interpretation was not thought necessary, and in-place interpretation, stymied by structured control flow, was dismissed as either impossible or at least unneeded.
Yet in the intervening five years, startup time and memory consumption have increased in priority, and interpretation of Wasm (by rewriting to an internal format) has become more widespread.

This paper restores the missing execution tier for Wasm, regaining the key property of efficient in-place interpretation that was thought lost.
We presented the design and implementation of the first fast in-place interpreter for Wasm that utilizes a compact sidetable easily generated as a side-effect of the code validation algorithm.
We measured an order of magnitude improvement in memory consumption and processing time over rewriting Wasm.
With this open problem now solved, we believe that Wasm engines in the future will employ new interpreter tiers for improved startup time, reduced memory consumption, and improved debugging support.

\begin{acks}
  Thanks to Paolo Severini at Microsoft for discussions on interpreter design and benchmarking considerations.
  Thanks to Hannes Payer and Toon Verwaest at Google for discussions on interpreters and V8.
  Thanks to Saam Barati, Keith Miller, and Fil Pizlo of the JavaScriptCore team for interpreter design discussions and tuning parameter hints.
  Thanks to Wasm Community Group members, specifically Andreas Rossberg, Luke Wagner, and Francis McCabe for discussions surrounding this work.
  Thanks to Lars T. Hansen at Mozilla for comments on a draft of this paper.
  Thanks to Tony Hosking and Steve Blackburn at ANU for shelter during (part of) the storm; truly Wizard is of Oz.
\end{acks}

\clearpage

\bibliography{paper}


\begin{thebibliography}{70}


\ifx \showCODEN    \undefined \def \showCODEN     #1{\unskip}     \fi
\ifx \showDOI      \undefined \def \showDOI       #1{#1}\fi
\ifx \showISBNx    \undefined \def \showISBNx     #1{\unskip}     \fi
\ifx \showISBNxiii \undefined \def \showISBNxiii  #1{\unskip}     \fi
\ifx \showISSN     \undefined \def \showISSN      #1{\unskip}     \fi
\ifx \showLCCN     \undefined \def \showLCCN      #1{\unskip}     \fi
\ifx \shownote     \undefined \def \shownote      #1{#1}          \fi
\ifx \showarticletitle \undefined \def \showarticletitle #1{#1}   \fi
\ifx \showURL      \undefined \def \showURL       {\relax}        \fi
\providecommand\bibfield[2]{#2}
\providecommand\bibinfo[2]{#2}
\providecommand\natexlab[1]{#1}
\providecommand\showeprint[2][]{arXiv:#2}

\bibitem[Hot(1998)]%
        {HotspotInt}
 \bibinfo{year}{1998}\natexlab{}.
\newblock \bibinfo{title}{Hotspot internals: interpreter}.
\newblock
  \bibinfo{howpublished}{\url{https://openjdk.java.net/groups/hotspot/docs/RuntimeOverview.html}}.
\newblock
\urldef\tempurl%
\url{https://openjdk.java.net/groups/hotspot/docs/RuntimeOverview.html}
\showURL{%
\tempurl}
\newblock
\shownote{(Accessed 2022-4-07)}.


\bibitem[Lli(2015)]%
        {Lli}
 \bibinfo{year}{2015}\natexlab{}.
\newblock \bibinfo{title}{lli - directly execute programs from LLVM bitcode}.
\newblock
  \bibinfo{howpublished}{\url{https://llvm.org/docs/CommandGuide/lli.html}}.
\newblock
\urldef\tempurl%
\url{https://llvm.org/docs/CommandGuide/lli.html}
\showURL{%
\tempurl}
\newblock
\shownote{(Accessed 2022-4-12)}.


\bibitem[Lif(2018)]%
        {Liftoff}
 \bibinfo{year}{2018}\natexlab{}.
\newblock \bibinfo{howpublished}{\url{https://v8.dev/blog/liftoff}}.
\newblock
\urldef\tempurl%
\url{https://v8.dev/blog/liftoff}
\showURL{%
\tempurl}
\newblock
\shownote{(Accessed 2022-4-07)}.


\bibitem[Tur(2018)]%
        {TurboFan}
 \bibinfo{year}{2018}\natexlab{}.
\newblock \bibinfo{title}{TurboFan: V8's Optimizing Compiler}.
\newblock \bibinfo{howpublished}{\url{https://v8.dev/docs/turbofan}}.
\newblock
\urldef\tempurl%
\url{https://v8.dev/docs/turbofan}
\showURL{%
\tempurl}
\newblock
\shownote{(Accessed 2021-07-29)}.


\bibitem[WAV(2018)]%
        {WAVM}
 \bibinfo{year}{2018}\natexlab{}.
\newblock \bibinfo{title}{{WAVM}: a non-browser {W}eb{A}ssembly virtual
  machine}.
\newblock \bibinfo{howpublished}{\url{https://github.com/WAVM/WAVM}}.
\newblock
\urldef\tempurl%
\url{https://github.com/WAVM/WAVM}
\showURL{%
\tempurl}
\newblock
\shownote{(Accessed 2022-1-10)}.


\bibitem[LLi(2019)]%
        {LLint}
 \bibinfo{year}{2019}\natexlab{}.
\newblock \bibinfo{title}{{A} {N}ew {B}ytecode {F}ormat for
  {J}ava{S}cript{C}ore}.
\newblock
  \bibinfo{howpublished}{\url{https://webkit.org/blog/9329/a-new-bytecode-format-for-javascriptcore/}}.
\newblock
\urldef\tempurl%
\url{https://webkit.org/blog/9329/a-new-bytecode-format-for-javascriptcore/}
\showURL{%
\tempurl}
\newblock
\shownote{(Accessed 2022-04-07)}.


\bibitem[Fas(2020)]%
        {FastlyEdge}
 \bibinfo{year}{2020}\natexlab{}.
\newblock \bibinfo{title}{The edge of the multi-cloud}.
\newblock
  \bibinfo{howpublished}{\url{https://www.fastly.com/cassets/6pk8mg3yh2ee/79dsHLTEfYIMgUwVVllaa4/5e5330572b8f317f72e16696256d8138/WhitePaper-Multi-Cloud.pdf}}.
\newblock
\urldef\tempurl%
\url{https://www.fastly.com/cassets/6pk8mg3yh2ee/79dsHLTEfYIMgUwVVllaa4/5e5330572b8f317f72e16696256d8138/WhitePaper-Multi-Cloud.pdf}
\showURL{%
\tempurl}
\newblock
\shownote{(Accessed 2021-07-06)}.


\bibitem[Was(2020)]%
        {Wasm3}
 \bibinfo{year}{2020}\natexlab{}.
\newblock \bibinfo{title}{{W}asm3: {T}he fastest {W}eb{A}ssembly interpreter,
  and the most universal runtime.}
\newblock \bibinfo{howpublished}{\url{https://github.com/wasm3/wasm3}}.
\newblock
\urldef\tempurl%
\url{https://github.com/wasm3/wasm3}
\showURL{%
\tempurl}
\newblock
\shownote{(Accessed 2021-08-11)}.


\bibitem[Jav(2021)]%
        {JavaScriptCore}
 \bibinfo{year}{2021}\natexlab{}.
\newblock \bibinfo{title}{Java{S}cript{C}ore, the built-in {J}ava{S}cript
  engine for {W}eb{K}it}.
\newblock
  \bibinfo{howpublished}{\url{https://trac.webkit.org/wiki/JavaScriptCore}}.
\newblock
\urldef\tempurl%
\url{https://trac.webkit.org/wiki/JavaScriptCore}
\showURL{%
\tempurl}
\newblock
\shownote{(Accessed 2021-07-29)}.


\bibitem[Spi(2021)]%
        {Spidermonkey}
 \bibinfo{year}{2021}\natexlab{}.
\newblock \bibinfo{title}{{S}pider{M}onkey: {M}ozilla’s {J}ava{S}cript and
  {W}eb{A}ssembly engine}.
\newblock \bibinfo{howpublished}{\url{https://spidermonkey.dev}}.
\newblock
\urldef\tempurl%
\url{https://spidermonkey.dev}
\showURL{%
\tempurl}
\newblock
\shownote{(Accessed 2021-07-29)}.


\bibitem[V8(2021)]%
        {V8}
 \bibinfo{year}{2021}\natexlab{}.
\newblock \bibinfo{title}{V8 Development Site}.
\newblock \bibinfo{howpublished}{\url{https://v8.dev}}.
\newblock
\urldef\tempurl%
\url{https://v8.dev}
\showURL{%
\tempurl}
\newblock
\shownote{(Accessed 2021-07-29)}.


\bibitem[Was(2021a)]%
        {Wasmer}
 \bibinfo{year}{2021}\natexlab{a}.
\newblock \bibinfo{title}{Wasmer: {A} {F}ast and {S}ecure {W}ebAssembly
  {R}untime}.
\newblock \bibinfo{howpublished}{\url{https://github.com/wasmerio/wasmer}}.
\newblock
\urldef\tempurl%
\url{https://github.com/wasmerio/wasmer}
\showURL{%
\tempurl}
\newblock
\shownote{(Accessed 2021-07-06)}.


\bibitem[Was(2021b)]%
        {Wasmtime}
 \bibinfo{year}{2021}\natexlab{b}.
\newblock \bibinfo{title}{{W}asmtime: a standalone runtime for
  {W}eb{A}ssembly}.
\newblock
  \bibinfo{howpublished}{\url{https://github.com/bytecodealliance/wasmtime}}.
\newblock
\urldef\tempurl%
\url{https://github.com/bytecodealliance/wasmtime}
\showURL{%
\tempurl}
\newblock
\shownote{(Accessed 2021-08-11)}.


\bibitem[Ign(2022)]%
        {Ignition}
 \bibinfo{year}{2022}\natexlab{}.
\newblock \bibinfo{title}{{I}gnition: a fast low-level interpreter}.
\newblock \bibinfo{howpublished}{\url{https://v8.dev/docs/ignition}}.
\newblock
\urldef\tempurl%
\url{https://v8.dev/docs/ignition}
\showURL{%
\tempurl}
\newblock
\shownote{(Accessed 2022-04-11)}.


\bibitem[Gcc(2022)]%
        {GccLabels}
 \bibinfo{year}{2022}\natexlab{}.
\newblock \bibinfo{title}{{L}abels as {V}alues ({GNU} {C}ompiler
  {C}ollection)}.
\newblock
  \bibinfo{howpublished}{\url{https://gcc.gnu.org/onlinedocs/gcc/Labels-as-Values.html}}.
\newblock
\urldef\tempurl%
\url{https://gcc.gnu.org/onlinedocs/gcc/Labels-as-Values.html}
\showURL{%
\tempurl}
\newblock
\shownote{(Accessed 2022-04-11)}.


\bibitem[LEB(2022)]%
        {LEB128}
 \bibinfo{year}{2022}\natexlab{}.
\newblock \bibinfo{title}{{LEB128}}.
\newblock \bibinfo{howpublished}{\url{https://en.wikipedia.org/wiki/LEB128}}.
\newblock
\urldef\tempurl%
\url{https://en.wikipedia.org/wiki/LEB128}
\showURL{%
\tempurl}
\newblock
\shownote{(Accessed 2022-04-11)}.


\bibitem[Wam(2022)]%
        {Wamr}
 \bibinfo{year}{2022}\natexlab{}.
\newblock \bibinfo{title}{{W}eb{A}ssembly {M}icro {R}untime ({WAMR})}.
\newblock
  \bibinfo{howpublished}{\url{https://github.com/bytecodealliance/wasm-micro-runtime}}.
\newblock
\urldef\tempurl%
\url{https://github.com/bytecodealliance/wasm-micro-runtime}
\showURL{%
\tempurl}
\newblock
\shownote{(Accessed 2022-04-11)}.


\bibitem[Amin and Rompf(2017)]%
        {CollapsingInt}
\bibfield{author}{\bibinfo{person}{Nada Amin} {and} \bibinfo{person}{Tiark
  Rompf}.} \bibinfo{year}{2017}\natexlab{}.
\newblock \showarticletitle{Collapsing Towers of Interpreters}.
\newblock \bibinfo{journal}{\emph{Proc. ACM Program. Lang.}}
  \bibinfo{volume}{2}, \bibinfo{number}{POPL}, Article \bibinfo{articleno}{52}
  (\bibinfo{date}{dec} \bibinfo{year}{2017}), \bibinfo{numpages}{33}~pages.
\newblock
\urldef\tempurl%
\url{https://doi.org/10.1145/3158140}
\showDOI{\tempurl}


\bibitem[Barany(2014)]%
        {PythonInt}
\bibfield{author}{\bibinfo{person}{Gerg\"{o} Barany}.}
  \bibinfo{year}{2014}\natexlab{}.
\newblock \showarticletitle{Python Interpreter Performance Deconstructed}. In
  \bibinfo{booktitle}{\emph{Proceedings of the Workshop on Dynamic Languages
  and Applications}} (Edinburgh, United Kingdom)
  \emph{(\bibinfo{series}{Dyla'14})}. \bibinfo{publisher}{Association for
  Computing Machinery}, \bibinfo{address}{New York, NY, USA},
  \bibinfo{pages}{1–9}.
\newblock
\showISBNx{9781450329163}
\urldef\tempurl%
\url{https://doi.org/10.1145/2617548.2617552}
\showDOI{\tempurl}


\bibitem[Bell(1973)]%
        {ThreadedCode}
\bibfield{author}{\bibinfo{person}{James~R. Bell}.}
  \bibinfo{year}{1973}\natexlab{}.
\newblock \showarticletitle{Threaded Code}.
\newblock \bibinfo{journal}{\emph{Commun. ACM}} \bibinfo{volume}{16},
  \bibinfo{number}{6} (\bibinfo{date}{jun} \bibinfo{year}{1973}),
  \bibinfo{pages}{370–372}.
\newblock
\showISSN{0001-0782}
\urldef\tempurl%
\url{https://doi.org/10.1145/362248.362270}
\showDOI{\tempurl}


\bibitem[Bolz et~al\mbox{.}(2009)]%
        {MetaTracing}
\bibfield{author}{\bibinfo{person}{Carl~Friedrich Bolz},
  \bibinfo{person}{Antonio Cuni}, \bibinfo{person}{Maciej Fijalkowski}, {and}
  \bibinfo{person}{Armin Rigo}.} \bibinfo{year}{2009}\natexlab{}.
\newblock \showarticletitle{Tracing the Meta-Level: PyPy's Tracing JIT
  Compiler}. In \bibinfo{booktitle}{\emph{Proceedings of the 4th Workshop on
  the Implementation, Compilation, Optimization of Object-Oriented Languages
  and Programming Systems}} (Genova, Italy) \emph{(\bibinfo{series}{ICOOOLPS
  '09})}. \bibinfo{publisher}{Association for Computing Machinery},
  \bibinfo{address}{New York, NY, USA}, \bibinfo{pages}{18–25}.
\newblock
\showISBNx{9781605585413}
\urldef\tempurl%
\url{https://doi.org/10.1145/1565824.1565827}
\showDOI{\tempurl}


\bibitem[Brunthaler(2010)]%
        {Quickening}
\bibfield{author}{\bibinfo{person}{Stefan Brunthaler}.}
  \bibinfo{year}{2010}\natexlab{}.
\newblock \showarticletitle{Efficient Interpretation Using Quickening}. In
  \bibinfo{booktitle}{\emph{Proceedings of the 6th Symposium on Dynamic
  Languages}} (Reno/Tahoe, Nevada, USA) \emph{(\bibinfo{series}{DLS '10})}.
  \bibinfo{publisher}{Association for Computing Machinery},
  \bibinfo{address}{New York, NY, USA}, \bibinfo{pages}{1–14}.
\newblock
\showISBNx{9781450304054}
\urldef\tempurl%
\url{https://doi.org/10.1145/1869631.1869633}
\showDOI{\tempurl}


\bibitem[Casey et~al\mbox{.}(2007)]%
        {OpIndirectAccuracy}
\bibfield{author}{\bibinfo{person}{Kevin Casey}, \bibinfo{person}{M.~Anton
  Ertl}, {and} \bibinfo{person}{David Gregg}.} \bibinfo{year}{2007}\natexlab{}.
\newblock \showarticletitle{Optimizing Indirect Branch Prediction Accuracy in
  Virtual Machine Interpreters}.
\newblock \bibinfo{journal}{\emph{ACM Trans. Program. Lang. Syst.}}
  \bibinfo{volume}{29}, \bibinfo{number}{6} (\bibinfo{date}{oct}
  \bibinfo{year}{2007}), \bibinfo{pages}{37–es}.
\newblock
\showISSN{0164-0925}
\urldef\tempurl%
\url{https://doi.org/10.1145/1286821.1286828}
\showDOI{\tempurl}


\bibitem[Casey et~al\mbox{.}(2005)]%
        {Tiger}
\bibfield{author}{\bibinfo{person}{Kevin Casey}, \bibinfo{person}{David Gregg},
  {and} \bibinfo{person}{M.~Anton Ertl}.} \bibinfo{year}{2005}\natexlab{}.
\newblock \showarticletitle{Tiger – an Interpreter Generation Tool}. In
  \bibinfo{booktitle}{\emph{Proceedings of the 14th International Conference on
  Compiler Construction}} (Edinburgh, UK) \emph{(\bibinfo{series}{CC'05})}.
  \bibinfo{publisher}{Springer-Verlag}, \bibinfo{address}{Berlin, Heidelberg},
  \bibinfo{pages}{246–249}.
\newblock
\showISBNx{3540254110}
\urldef\tempurl%
\url{https://doi.org/10.1007/978-3-540-31985-6_18}
\showDOI{\tempurl}


\bibitem[Chang et~al\mbox{.}(1997)]%
        {BTBs}
\bibfield{author}{\bibinfo{person}{Po-Yung Chang}, \bibinfo{person}{Eric Hao},
  {and} \bibinfo{person}{Yale~N Patt}.} \bibinfo{year}{1997}\natexlab{}.
\newblock \showarticletitle{Target prediction for indirect jumps}.
\newblock \bibinfo{journal}{\emph{ACM SIGARCH Computer Architecture News}}
  \bibinfo{volume}{25}, \bibinfo{number}{2} (\bibinfo{year}{1997}),
  \bibinfo{pages}{274--283}.
\newblock


\bibitem[Curtsinger and Berger(2013)]%
        {Stabilizer}
\bibfield{author}{\bibinfo{person}{Charlie Curtsinger} {and}
  \bibinfo{person}{Emery~D. Berger}.} \bibinfo{year}{2013}\natexlab{}.
\newblock \showarticletitle{STABILIZER: Statistically Sound Performance
  Evaluation}. In \bibinfo{booktitle}{\emph{Proceedings of the Eighteenth
  International Conference on Architectural Support for Programming Languages
  and Operating Systems}} (Houston, Texas, USA) \emph{(\bibinfo{series}{ASPLOS
  '13})}. \bibinfo{publisher}{Association for Computing Machinery},
  \bibinfo{address}{New York, NY, USA}, \bibinfo{pages}{219–228}.
\newblock
\showISBNx{9781450318709}
\urldef\tempurl%
\url{https://doi.org/10.1145/2451116.2451141}
\showDOI{\tempurl}


\bibitem[Davis et~al\mbox{.}(2003)]%
        {VirtRegMachine}
\bibfield{author}{\bibinfo{person}{Brian Davis}, \bibinfo{person}{Andrew
  Beatty}, \bibinfo{person}{Kevin Casey}, \bibinfo{person}{David Gregg}, {and}
  \bibinfo{person}{John Waldron}.} \bibinfo{year}{2003}\natexlab{}.
\newblock \showarticletitle{The Case for Virtual Register Machines}. In
  \bibinfo{booktitle}{\emph{Proceedings of the 2003 Workshop on Interpreters,
  Virtual Machines and Emulators}} (San Diego, California)
  \emph{(\bibinfo{series}{IVME '03})}. \bibinfo{publisher}{Association for
  Computing Machinery}, \bibinfo{address}{New York, NY, USA},
  \bibinfo{pages}{41–49}.
\newblock
\showISBNx{1581136552}
\urldef\tempurl%
\url{https://doi.org/10.1145/858570.858575}
\showDOI{\tempurl}


\bibitem[Demers et~al\mbox{.}(1989)]%
        {ConservativeGc}
\bibfield{author}{\bibinfo{person}{Alan Demers}, \bibinfo{person}{Mark Weiser},
  \bibinfo{person}{Barry Hayes}, \bibinfo{person}{Hans Boehm},
  \bibinfo{person}{Daniel Bobrow}, {and} \bibinfo{person}{Scott Shenker}.}
  \bibinfo{year}{1989}\natexlab{}.
\newblock \showarticletitle{Combining Generational and Conservative Garbage
  Collection: Framework and Implementations}. In
  \bibinfo{booktitle}{\emph{Proceedings of the 17th ACM SIGPLAN-SIGACT
  Symposium on Principles of Programming Languages}} (San Francisco,
  California, USA) \emph{(\bibinfo{series}{POPL '90})}.
  \bibinfo{publisher}{Association for Computing Machinery},
  \bibinfo{address}{New York, NY, USA}, \bibinfo{pages}{261–269}.
\newblock
\showISBNx{0897913434}
\urldef\tempurl%
\url{https://doi.org/10.1145/96709.96735}
\showDOI{\tempurl}


\bibitem[Dewar(1975)]%
        {IndirectThreadedCode}
\bibfield{author}{\bibinfo{person}{Robert B.~K. Dewar}.}
  \bibinfo{year}{1975}\natexlab{}.
\newblock \showarticletitle{Indirect Threaded Code}.
\newblock \bibinfo{journal}{\emph{Commun. ACM}} \bibinfo{volume}{18},
  \bibinfo{number}{6} (\bibinfo{date}{jun} \bibinfo{year}{1975}),
  \bibinfo{pages}{330–331}.
\newblock
\showISSN{0001-0782}
\urldef\tempurl%
\url{https://doi.org/10.1145/360825.360849}
\showDOI{\tempurl}


\bibitem[Diwan et~al\mbox{.}(1992)]%
        {GcStackmaps}
\bibfield{author}{\bibinfo{person}{Amer Diwan}, \bibinfo{person}{Eliot Moss},
  {and} \bibinfo{person}{Richard Hudson}.} \bibinfo{year}{1992}\natexlab{}.
\newblock \showarticletitle{Compiler Support for Garbage Collection in a
  Statically Typed Language}. In \bibinfo{booktitle}{\emph{Proceedings of the
  ACM SIGPLAN 1992 Conference on Programming Language Design and
  Implementation}} (San Francisco, California, USA)
  \emph{(\bibinfo{series}{PLDI '92})}. \bibinfo{publisher}{Association for
  Computing Machinery}, \bibinfo{address}{New York, NY, USA},
  \bibinfo{pages}{273–282}.
\newblock
\showISBNx{0897914759}
\urldef\tempurl%
\url{https://doi.org/10.1145/143095.143140}
\showDOI{\tempurl}


\bibitem[Ertl and Gregg(2004a)]%
        {DynamicSuperInst}
\bibfield{author}{\bibinfo{person}{M.~Anton Ertl} {and} \bibinfo{person}{David
  Gregg}.} \bibinfo{year}{2004}\natexlab{a}.
\newblock \showarticletitle{Combining Stack Caching with Dynamic
  Superinstructions}. In \bibinfo{booktitle}{\emph{Proceedings of the 2004
  Workshop on Interpreters, Virtual Machines and Emulators}} (Washington, D.C.)
  \emph{(\bibinfo{series}{IVME '04})}. \bibinfo{publisher}{Association for
  Computing Machinery}, \bibinfo{address}{New York, NY, USA},
  \bibinfo{pages}{7–14}.
\newblock
\showISBNx{1581139098}
\urldef\tempurl%
\url{https://doi.org/10.1145/1059579.1059583}
\showDOI{\tempurl}


\bibitem[Ertl and Gregg(2004b)]%
        {CombiningStackCaching}
\bibfield{author}{\bibinfo{person}{M.~Anton Ertl} {and} \bibinfo{person}{David
  Gregg}.} \bibinfo{year}{2004}\natexlab{b}.
\newblock \showarticletitle{Combining Stack Caching with Dynamic
  Superinstructions}. In \bibinfo{booktitle}{\emph{Proceedings of the 2004
  Workshop on Interpreters, Virtual Machines and Emulators}} (Washington, D.C.)
  \emph{(\bibinfo{series}{IVME '04})}. \bibinfo{publisher}{Association for
  Computing Machinery}, \bibinfo{address}{New York, NY, USA},
  \bibinfo{pages}{7–14}.
\newblock
\showISBNx{1581139098}
\urldef\tempurl%
\url{https://doi.org/10.1145/1059579.1059583}
\showDOI{\tempurl}


\bibitem[Ertl et~al\mbox{.}(2002)]%
        {vmgen}
\bibfield{author}{\bibinfo{person}{M.~Anton Ertl}, \bibinfo{person}{David
  Gregg}, \bibinfo{person}{Andreas Krall}, {and} \bibinfo{person}{Bernd
  Paysan}.} \bibinfo{year}{2002}\natexlab{}.
\newblock \showarticletitle{Vmgen: A Generator of Efficient Virtual Machine
  Interpreters}.
\newblock \bibinfo{journal}{\emph{Softw. Pract. Exper.}} \bibinfo{volume}{32},
  \bibinfo{number}{3} (\bibinfo{date}{mar} \bibinfo{year}{2002}),
  \bibinfo{pages}{265–294}.
\newblock
\showISSN{0038-0644}
\urldef\tempurl%
\url{https://doi.org/10.1002/spe.434}
\showDOI{\tempurl}


\bibitem[Futamura(1983)]%
        {Futamura}
\bibfield{author}{\bibinfo{person}{Y. Futamura}.}
  \bibinfo{year}{1983}\natexlab{}.
\newblock \showarticletitle{Partial computation of programs.}
\newblock \bibinfo{journal}{\emph{Lecture Notes in Computer Science}}
  \bibinfo{volume}{147} (\bibinfo{year}{1983}).
\newblock


\bibitem[Gal et~al\mbox{.}(2009)]%
        {TraceCompilation}
\bibfield{author}{\bibinfo{person}{Andreas Gal}, \bibinfo{person}{Brendan
  Eich}, \bibinfo{person}{Mike Shaver}, \bibinfo{person}{David Anderson},
  \bibinfo{person}{David Mandelin}, \bibinfo{person}{Mohammad~R. Haghighat},
  \bibinfo{person}{Blake Kaplan}, \bibinfo{person}{Graydon Hoare},
  \bibinfo{person}{Boris Zbarsky}, \bibinfo{person}{Jason Orendorff},
  \bibinfo{person}{Jesse Ruderman}, \bibinfo{person}{Edwin~W. Smith},
  \bibinfo{person}{Rick Reitmaier}, \bibinfo{person}{Michael Bebenita},
  \bibinfo{person}{Mason Chang}, {and} \bibinfo{person}{Michael Franz}.}
  \bibinfo{year}{2009}\natexlab{}.
\newblock \showarticletitle{Trace-Based Just-in-Time Type Specialization for
  Dynamic Languages}. In \bibinfo{booktitle}{\emph{Proceedings of the 30th ACM
  SIGPLAN Conference on Programming Language Design and Implementation}}
  (Dublin, Ireland) \emph{(\bibinfo{series}{PLDI '09})}.
  \bibinfo{publisher}{Association for Computing Machinery},
  \bibinfo{address}{New York, NY, USA}, \bibinfo{pages}{465–478}.
\newblock
\showISBNx{9781605583921}
\urldef\tempurl%
\url{https://doi.org/10.1145/1542476.1542528}
\showDOI{\tempurl}


\bibitem[Gurdeep~Singh and Scholliers(2019)]%
        {WarDuino}
\bibfield{author}{\bibinfo{person}{Robbert Gurdeep~Singh} {and}
  \bibinfo{person}{Christophe Scholliers}.} \bibinfo{year}{2019}\natexlab{}.
\newblock \showarticletitle{WARDuino: A Dynamic {W}eb{A}ssembly Virtual Machine
  for Programming Microcontrollers}. In \bibinfo{booktitle}{\emph{Proceedings
  of the 16th ACM SIGPLAN International Conference on Managed Programming
  Languages and Runtimes}} (Athens, Greece) \emph{(\bibinfo{series}{MPLR
  2019})}. \bibinfo{publisher}{Association for Computing Machinery},
  \bibinfo{address}{New York, NY, USA}, \bibinfo{pages}{27–36}.
\newblock
\showISBNx{9781450369770}
\urldef\tempurl%
\url{https://doi.org/10.1145/3357390.3361029}
\showDOI{\tempurl}


\bibitem[Haas et~al\mbox{.}(2017)]%
        {WasmPldi}
\bibfield{author}{\bibinfo{person}{Andreas Haas}, \bibinfo{person}{Andreas
  Rossberg}, \bibinfo{person}{Derek~L. Schuff}, \bibinfo{person}{Ben~L.
  Titzer}, \bibinfo{person}{Michael Holman}, \bibinfo{person}{Dan Gohman},
  \bibinfo{person}{Luke Wagner}, \bibinfo{person}{Alon Zakai}, {and}
  \bibinfo{person}{JF Bastien}.} \bibinfo{year}{2017}\natexlab{}.
\newblock \showarticletitle{Bringing the Web up to Speed with {W}eb{A}ssembly}.
  In \bibinfo{booktitle}{\emph{Proceedings of the 38th ACM SIGPLAN Conference
  on Programming Language Design and Implementation}} (Barcelona, Spain)
  \emph{(\bibinfo{series}{PLDI 2017})}. \bibinfo{publisher}{Association for
  Computing Machinery}, \bibinfo{address}{New York, NY, USA},
  \bibinfo{pages}{185–200}.
\newblock
\showISBNx{9781450349888}
\urldef\tempurl%
\url{https://doi.org/10.1145/3062341.3062363}
\showDOI{\tempurl}


\bibitem[Hu et~al\mbox{.}(2021)]%
        {DatalogInterp}
\bibfield{author}{\bibinfo{person}{Xiaowen Hu}, \bibinfo{person}{David Zhao},
  \bibinfo{person}{Herbert Jordan}, {and} \bibinfo{person}{Bernhard Scholz}.}
  \bibinfo{year}{2021}\natexlab{}.
\newblock \showarticletitle{An Efficient Interpreter for Datalog by
  De-Specializing Relations} \emph{(\bibinfo{series}{PLDI 2021})}.
  \bibinfo{publisher}{Association for Computing Machinery},
  \bibinfo{address}{New York, NY, USA}, \bibinfo{pages}{681–695}.
\newblock
\showISBNx{9781450383912}
\urldef\tempurl%
\url{https://doi.org/10.1145/3453483.3454070}
\showDOI{\tempurl}


\bibitem[Kalibera et~al\mbox{.}(2014)]%
        {RAstInt}
\bibfield{author}{\bibinfo{person}{Tomas Kalibera}, \bibinfo{person}{Petr Maj},
  \bibinfo{person}{Floreal Morandat}, {and} \bibinfo{person}{Jan Vitek}.}
  \bibinfo{year}{2014}\natexlab{}.
\newblock \showarticletitle{A Fast Abstract Syntax Tree Interpreter for R}. In
  \bibinfo{booktitle}{\emph{Proceedings of the 10th ACM SIGPLAN/SIGOPS
  International Conference on Virtual Execution Environments}} (Salt Lake City,
  Utah, USA) \emph{(\bibinfo{series}{VEE '14})}.
  \bibinfo{publisher}{Association for Computing Machinery},
  \bibinfo{address}{New York, NY, USA}, \bibinfo{pages}{89–102}.
\newblock
\showISBNx{9781450327640}
\urldef\tempurl%
\url{https://doi.org/10.1145/2576195.2576205}
\showDOI{\tempurl}


\bibitem[Kataoka et~al\mbox{.}(2018)]%
        {JsvmGen}
\bibfield{author}{\bibinfo{person}{Takafumi Kataoka}, \bibinfo{person}{Tomoharu
  Ugawa}, {and} \bibinfo{person}{Hideya Iwasaki}.}
  \bibinfo{year}{2018}\natexlab{}.
\newblock \showarticletitle{A Framework for Constructing Javascript Virtual
  Machines with Customized Datatype Representations}. In
  \bibinfo{booktitle}{\emph{Proceedings of the 33rd Annual ACM Symposium on
  Applied Computing}} (Pau, France) \emph{(\bibinfo{series}{SAC '18})}.
  \bibinfo{publisher}{Association for Computing Machinery},
  \bibinfo{address}{New York, NY, USA}, \bibinfo{pages}{1238–1247}.
\newblock
\showISBNx{9781450351911}
\urldef\tempurl%
\url{https://doi.org/10.1145/3167132.3167266}
\showDOI{\tempurl}


\bibitem[Kim et~al\mbox{.}(2016)]%
        {ShortCircuitDispatch}
\bibfield{author}{\bibinfo{person}{Channoh Kim}, \bibinfo{person}{Sungmin Kim},
  \bibinfo{person}{Hyeon~Gyu Cho}, \bibinfo{person}{Dooyoung Kim},
  \bibinfo{person}{Jaehyeok Kim}, \bibinfo{person}{Young~H. Oh},
  \bibinfo{person}{Hakbeom Jang}, {and} \bibinfo{person}{Jae~W. Lee}.}
  \bibinfo{year}{2016}\natexlab{}.
\newblock \showarticletitle{Short-Circuit Dispatch: Accelerating Virtual
  Machine Interpreters on Embedded Processors}. In
  \bibinfo{booktitle}{\emph{2016 ACM/IEEE 43rd Annual International Symposium
  on Computer Architecture (ISCA)}}. \bibinfo{pages}{291--303}.
\newblock
\urldef\tempurl%
\url{https://doi.org/10.1109/ISCA.2016.34}
\showDOI{\tempurl}


\bibitem[Li et~al\mbox{.}(2021)]%
        {WasmIotOs}
\bibfield{author}{\bibinfo{person}{Borui Li}, \bibinfo{person}{Hongchang Fan},
  \bibinfo{person}{Yi Gao}, {and} \bibinfo{person}{Wei Dong}.}
  \bibinfo{year}{2021}\natexlab{}.
\newblock \showarticletitle{ThingSpire OS: A {W}eb{A}ssembly-Based IoT
  Operating System for Cloud-Edge Integration}. In
  \bibinfo{booktitle}{\emph{Proceedings of the 19th Annual International
  Conference on Mobile Systems, Applications, and Services}} (Virtual Event,
  Wisconsin) \emph{(\bibinfo{series}{MobiSys '21})}.
  \bibinfo{publisher}{Association for Computing Machinery},
  \bibinfo{address}{New York, NY, USA}, \bibinfo{pages}{487–488}.
\newblock
\showISBNx{9781450384438}
\urldef\tempurl%
\url{https://doi.org/10.1145/3458864.3466910}
\showDOI{\tempurl}


\bibitem[McCarthy(1960)]%
        {Lisp}
\bibfield{author}{\bibinfo{person}{John McCarthy}.}
  \bibinfo{year}{1960}\natexlab{}.
\newblock \showarticletitle{Recursive Functions of Symbolic Expressions and
  Their Computation by Machine, Part I}.
\newblock \bibinfo{journal}{\emph{Commun. ACM}} \bibinfo{volume}{3},
  \bibinfo{number}{4} (\bibinfo{date}{apr} \bibinfo{year}{1960}),
  \bibinfo{pages}{184–195}.
\newblock
\showISSN{0001-0782}
\urldef\tempurl%
\url{https://doi.org/10.1145/367177.367199}
\showDOI{\tempurl}


\bibitem[Microsoft(2021)]%
        {ChakraCore}
\bibfield{author}{\bibinfo{person}{Microsoft}.}
  \bibinfo{year}{2021}\natexlab{}.
\newblock \bibinfo{title}{{C}hakra{C}ore: a {J}ava{S}cript engine with a C
  API}.
\newblock
  \bibinfo{howpublished}{\url{https://github.com/chakra-core/ChakraCore}}.
\newblock
\urldef\tempurl%
\url{https://github.com/chakra-core/ChakraCore}
\showURL{%
\tempurl}
\newblock
\shownote{(Accessed 2021-07-29)}.


\bibitem[Mytkowicz et~al\mbox{.}(2009)]%
        {WrongData}
\bibfield{author}{\bibinfo{person}{Todd Mytkowicz}, \bibinfo{person}{Amer
  Diwan}, \bibinfo{person}{Matthias Hauswirth}, {and} \bibinfo{person}{Peter~F.
  Sweeney}.} \bibinfo{year}{2009}\natexlab{}.
\newblock \showarticletitle{Producing Wrong Data without Doing Anything
  Obviously Wrong!}. In \bibinfo{booktitle}{\emph{Proceedings of the 14th
  International Conference on Architectural Support for Programming Languages
  and Operating Systems}} (Washington, DC, USA) \emph{(\bibinfo{series}{ASPLOS
  XIV})}. \bibinfo{publisher}{Association for Computing Machinery},
  \bibinfo{address}{New York, NY, USA}, \bibinfo{pages}{265–276}.
\newblock
\showISBNx{9781605584065}
\urldef\tempurl%
\url{https://doi.org/10.1145/1508244.1508275}
\showDOI{\tempurl}


\bibitem[Nieke et~al\mbox{.}(2021)]%
        {WasmEdgeDancer}
\bibfield{author}{\bibinfo{person}{Manuel Nieke}, \bibinfo{person}{Lennart
  Almstedt}, {and} \bibinfo{person}{R\"{u}diger Kapitza}.}
  \bibinfo{year}{2021}\natexlab{}.
\newblock \showarticletitle{Edgedancer: Secure Mobile {W}eb{A}ssembly Services
  on the Edge}. In \bibinfo{booktitle}{\emph{Proceedings of the 4th
  International Workshop on Edge Systems, Analytics and Networking}} (Online,
  United Kingdom) \emph{(\bibinfo{series}{EdgeSys '21})}.
  \bibinfo{publisher}{Association for Computing Machinery},
  \bibinfo{address}{New York, NY, USA}, \bibinfo{pages}{13–18}.
\newblock
\showISBNx{9781450382915}
\urldef\tempurl%
\url{https://doi.org/10.1145/3434770.3459731}
\showDOI{\tempurl}


\bibitem[Niephaus et~al\mbox{.}(2018)]%
        {GraalSqueak}
\bibfield{author}{\bibinfo{person}{Fabio Niephaus}, \bibinfo{person}{Tim
  Felgentreff}, {and} \bibinfo{person}{Robert Hirschfeld}.}
  \bibinfo{year}{2018}\natexlab{}.
\newblock \showarticletitle{GraalSqueak: A Fast Smalltalk Bytecode Interpreter
  Written in an AST Interpreter Framework}. In
  \bibinfo{booktitle}{\emph{Proceedings of the 13th Workshop on Implementation,
  Compilation, Optimization of Object-Oriented Languages, Programs and
  Systems}} (Amsterdam, Netherlands) \emph{(\bibinfo{series}{ICOOOLPS '18})}.
  \bibinfo{publisher}{Association for Computing Machinery},
  \bibinfo{address}{New York, NY, USA}, \bibinfo{pages}{30–35}.
\newblock
\showISBNx{9781450358040}
\urldef\tempurl%
\url{https://doi.org/10.1145/3242947.3242948}
\showDOI{\tempurl}


\bibitem[Nurul-Hoque and Harras(2021)]%
        {FemtoCloud}
\bibfield{author}{\bibinfo{person}{Mohammed Nurul-Hoque} {and}
  \bibinfo{person}{Khaled~A. Harras}.} \bibinfo{year}{2021}\natexlab{}.
\newblock \showarticletitle{Nomad: Cross-Platform Computational Offloading and
  Migration in Femtoclouds Using WebAssembly}. In
  \bibinfo{booktitle}{\emph{2021 IEEE International Conference on Cloud
  Engineering (IC2E)}}. \bibinfo{pages}{168--178}.
\newblock
\urldef\tempurl%
\url{https://doi.org/10.1109/IC2E52221.2021.00032}
\showDOI{\tempurl}


\bibitem[Ogata et~al\mbox{.}(2002)]%
        {JavaIntOpt}
\bibfield{author}{\bibinfo{person}{Kazunori Ogata}, \bibinfo{person}{Hideaki
  Komatsu}, {and} \bibinfo{person}{Toshio Nakatani}.}
  \bibinfo{year}{2002}\natexlab{}.
\newblock \showarticletitle{Bytecode Fetch Optimization for a Java
  Interpreter}. In \bibinfo{booktitle}{\emph{Proceedings of the 10th
  International Conference on Architectural Support for Programming Languages
  and Operating Systems}} (San Jose, California) \emph{(\bibinfo{series}{ASPLOS
  X})}. \bibinfo{publisher}{Association for Computing Machinery},
  \bibinfo{address}{New York, NY, USA}, \bibinfo{pages}{58–67}.
\newblock
\showISBNx{1581135742}
\urldef\tempurl%
\url{https://doi.org/10.1145/605397.605404}
\showDOI{\tempurl}


\bibitem[Palsberg(2014)]%
        {Cs132}
\bibfield{author}{\bibinfo{person}{Jens Palsberg}.}
  \bibinfo{year}{2014}\natexlab{}.
\newblock \bibinfo{title}{From Java to Mips in Four Nifty Steps}.
\newblock
  \bibinfo{howpublished}{\url{http://web.cs.ucla.edu/classes/spring11/cs132/kannan/index.html}}.
\newblock
\urldef\tempurl%
\url{http://web.cs.ucla.edu/classes/spring11/cs132/kannan/index.html}
\showURL{%
\tempurl}
\newblock
\shownote{(Accessed 2022-4-07)}.


\bibitem[Peng et~al\mbox{.}(2004)]%
        {StackCaching}
\bibfield{author}{\bibinfo{person}{Jinzhan Peng}, \bibinfo{person}{Gansha Wu},
  {and} \bibinfo{person}{Guei-Yuan Lueh}.} \bibinfo{year}{2004}\natexlab{}.
\newblock \showarticletitle{Code Sharing among States for Stack-Caching
  Interpreter}. In \bibinfo{booktitle}{\emph{Proceedings of the 2004 Workshop
  on Interpreters, Virtual Machines and Emulators}} (Washington, D.C.)
  \emph{(\bibinfo{series}{IVME '04})}. \bibinfo{publisher}{Association for
  Computing Machinery}, \bibinfo{address}{New York, NY, USA},
  \bibinfo{pages}{15–22}.
\newblock
\showISBNx{1581139098}
\urldef\tempurl%
\url{https://doi.org/10.1145/1059579.1059584}
\showDOI{\tempurl}


\bibitem[Piumarta and Riccardi(1998)]%
        {OptDirectThreadInline}
\bibfield{author}{\bibinfo{person}{Ian Piumarta} {and} \bibinfo{person}{Fabio
  Riccardi}.} \bibinfo{year}{1998}\natexlab{}.
\newblock \showarticletitle{Optimizing Direct Threaded Code by Selective
  Inlining}. In \bibinfo{booktitle}{\emph{Proceedings of the ACM SIGPLAN 1998
  Conference on Programming Language Design and Implementation}} (Montreal,
  Quebec, Canada) \emph{(\bibinfo{series}{PLDI '98})}.
  \bibinfo{publisher}{Association for Computing Machinery},
  \bibinfo{address}{New York, NY, USA}, \bibinfo{pages}{291–300}.
\newblock
\showISBNx{0897919874}
\urldef\tempurl%
\url{https://doi.org/10.1145/277650.277743}
\showDOI{\tempurl}


\bibitem[Pizlo(2016)]%
        {B3}
\bibfield{author}{\bibinfo{person}{Filip Pizlo}.}
  \bibinfo{year}{2016}\natexlab{}.
\newblock \bibinfo{title}{Introducing the B3 JIT Compiler}.
\newblock
  \bibinfo{howpublished}{\url{https://webkit.org/blog/5852/introducing-the-b3-jit-compiler/}}.
\newblock
\urldef\tempurl%
\url{https://webkit.org/blog/5852/introducing-the-b3-jit-compiler/}
\showURL{%
\tempurl}
\newblock
\shownote{(Accessed 2022-4-12)}.


\bibitem[Proebsting(1995)]%
        {Superinstructions}
\bibfield{author}{\bibinfo{person}{Todd~A. Proebsting}.}
  \bibinfo{year}{1995}\natexlab{}.
\newblock \showarticletitle{Optimizing an ANSI C Interpreter with
  Superoperators}. In \bibinfo{booktitle}{\emph{Proceedings of the 22nd ACM
  SIGPLAN-SIGACT Symposium on Principles of Programming Languages}} (San
  Francisco, California, USA) \emph{(\bibinfo{series}{POPL '95})}.
  \bibinfo{publisher}{Association for Computing Machinery},
  \bibinfo{address}{New York, NY, USA}, \bibinfo{pages}{322–332}.
\newblock
\showISBNx{0897916921}
\urldef\tempurl%
\url{https://doi.org/10.1145/199448.199526}
\showDOI{\tempurl}


\bibitem[Prokopski and Verbrugge(2008)]%
        {CodeCopyingVM}
\bibfield{author}{\bibinfo{person}{Gregory~B. Prokopski} {and}
  \bibinfo{person}{Clark Verbrugge}.} \bibinfo{year}{2008}\natexlab{}.
\newblock \showarticletitle{Analyzing the Performance of Code-Copying Virtual
  Machines}.
\newblock \bibinfo{journal}{\emph{SIGPLAN Not.}} \bibinfo{volume}{43},
  \bibinfo{number}{10} (\bibinfo{date}{oct} \bibinfo{year}{2008}),
  \bibinfo{pages}{403–422}.
\newblock
\showISSN{0362-1340}
\urldef\tempurl%
\url{https://doi.org/10.1145/1449955.1449796}
\showDOI{\tempurl}


\bibitem[Rohou et~al\mbox{.}(2015)]%
        {BranchPredInt}
\bibfield{author}{\bibinfo{person}{Erven Rohou},
  \bibinfo{person}{Bharath~Narasimha Swamy}, {and} \bibinfo{person}{André
  Seznec}.} \bibinfo{year}{2015}\natexlab{}.
\newblock \showarticletitle{Branch prediction and the performance of
  interpreters — Don't trust folklore}. In \bibinfo{booktitle}{\emph{2015
  IEEE/ACM International Symposium on Code Generation and Optimization (CGO)}}.
  \bibinfo{pages}{103--114}.
\newblock
\urldef\tempurl%
\url{https://doi.org/10.1109/CGO.2015.7054191}
\showDOI{\tempurl}


\bibitem[Savrun-Yeni\c{c}eri et~al\mbox{.}(2014)]%
        {HostedInterpreters}
\bibfield{author}{\bibinfo{person}{G\"{u}lfem Savrun-Yeni\c{c}eri},
  \bibinfo{person}{Wei Zhang}, \bibinfo{person}{Huahan Zhang},
  \bibinfo{person}{Eric Seckler}, \bibinfo{person}{Chen Li},
  \bibinfo{person}{Stefan Brunthaler}, \bibinfo{person}{Per Larsen}, {and}
  \bibinfo{person}{Michael Franz}.} \bibinfo{year}{2014}\natexlab{}.
\newblock \showarticletitle{Efficient Hosted Interpreters on the JVM}.
\newblock \bibinfo{journal}{\emph{ACM Trans. Archit. Code Optim.}}
  \bibinfo{volume}{11}, \bibinfo{number}{1}, Article \bibinfo{articleno}{9}
  (\bibinfo{date}{feb} \bibinfo{year}{2014}), \bibinfo{numpages}{24}~pages.
\newblock
\showISSN{1544-3566}
\urldef\tempurl%
\url{https://doi.org/10.1145/2532642}
\showDOI{\tempurl}


\bibitem[Shi et~al\mbox{.}(2005)]%
        {VmShowdown}
\bibfield{author}{\bibinfo{person}{Yunhe Shi}, \bibinfo{person}{David Gregg},
  \bibinfo{person}{Andrew Beatty}, {and} \bibinfo{person}{M.~Anton Ertl}.}
  \bibinfo{year}{2005}\natexlab{}.
\newblock \showarticletitle{{V}irtual {M}achine {S}howdown: {S}tack versus
  {R}egisters}. In \bibinfo{booktitle}{\emph{Proceedings of the 1st ACM/USENIX
  International Conference on Virtual Execution Environments}} (Chicago, IL,
  USA) \emph{(\bibinfo{series}{VEE '05})}. \bibinfo{publisher}{Association for
  Computing Machinery}, \bibinfo{address}{New York, NY, USA},
  \bibinfo{pages}{153–163}.
\newblock
\showISBNx{1595930477}
\urldef\tempurl%
\url{https://doi.org/10.1145/1064979.1065001}
\showDOI{\tempurl}


\bibitem[Titzer(2013)]%
        {VirgilPldi}
\bibfield{author}{\bibinfo{person}{Ben~L. Titzer}.}
  \bibinfo{year}{2013}\natexlab{}.
\newblock \showarticletitle{Harmonizing Classes, Functions, Tuples, and Type
  Parameters in Virgil {III}}. In \bibinfo{booktitle}{\emph{Proceedings of the
  34th ACM SIGPLAN Conference on Programming Language Design and
  Implementation}} (Seattle, Washington, USA) \emph{(\bibinfo{series}{PLDI
  '13})}. \bibinfo{publisher}{Association for Computing Machinery},
  \bibinfo{address}{New York, NY, USA}, \bibinfo{pages}{85–94}.
\newblock
\showISBNx{9781450320146}
\urldef\tempurl%
\url{https://doi.org/10.1145/2491956.2491962}
\showDOI{\tempurl}


\bibitem[Titzer(2021)]%
        {WizardEngine}
\bibfield{author}{\bibinfo{person}{Ben~L. Titzer}.}
  \bibinfo{year}{2021}\natexlab{}.
\newblock \bibinfo{title}{{W}izard, {A}n advanced {W}ebAssembly {E}ngine for
  {R}esearch}.
\newblock
  \bibinfo{howpublished}{\url{https://github.com/titzer/wizard-engine}}.
\newblock
\urldef\tempurl%
\url{https://github.com/titzer/wizard-engine}
\showURL{%
\tempurl}
\newblock
\shownote{(Accessed 2021-07-29)}.


\bibitem[Titzer and Palsberg(2005)]%
        {NonIntrusive}
\bibfield{author}{\bibinfo{person}{Ben~L. Titzer} {and} \bibinfo{person}{Jens
  Palsberg}.} \bibinfo{year}{2005}\natexlab{}.
\newblock \showarticletitle{Nonintrusive Precision Instrumentation of
  Microcontroller Software}. In \bibinfo{booktitle}{\emph{Proceedings of the
  2005 ACM SIGPLAN/SIGBED Conference on Languages, Compilers, and Tools for
  Embedded Systems}} (Chicago, Illinois, USA) \emph{(\bibinfo{series}{LCTES
  '05})}. \bibinfo{publisher}{Association for Computing Machinery},
  \bibinfo{address}{New York, NY, USA}, \bibinfo{pages}{59–68}.
\newblock
\showISBNx{1595930183}
\urldef\tempurl%
\url{https://doi.org/10.1145/1065910.1065919}
\showDOI{\tempurl}


\bibitem[Williams et~al\mbox{.}(2010)]%
        {DynamicInterp}
\bibfield{author}{\bibinfo{person}{Kevin Williams}, \bibinfo{person}{Jason
  McCandless}, {and} \bibinfo{person}{David Gregg}.}
  \bibinfo{year}{2010}\natexlab{}.
\newblock \showarticletitle{Dynamic Interpretation for Dynamic Scripting
  Languages}. In \bibinfo{booktitle}{\emph{Proceedings of the 8th Annual
  IEEE/ACM International Symposium on Code Generation and Optimization}}
  (Toronto, Ontario, Canada) \emph{(\bibinfo{series}{CGO '10})}.
  \bibinfo{publisher}{Association for Computing Machinery},
  \bibinfo{address}{New York, NY, USA}, \bibinfo{pages}{278–287}.
\newblock
\showISBNx{9781605586359}
\urldef\tempurl%
\url{https://doi.org/10.1145/1772954.1772993}
\showDOI{\tempurl}


\bibitem[Williams et~al\mbox{.}(2008)]%
        {JavaOnCell}
\bibfield{author}{\bibinfo{person}{Kevin Williams}, \bibinfo{person}{Albert
  Noll}, \bibinfo{person}{Andreas Gal}, {and} \bibinfo{person}{David Gregg}.}
  \bibinfo{year}{2008}\natexlab{}.
\newblock \showarticletitle{Optimization Strategies for a Java Virtual Machine
  Interpreter on the Cell Broadband Engine}. In
  \bibinfo{booktitle}{\emph{Proceedings of the 5th Conference on Computing
  Frontiers}} (Ischia, Italy) \emph{(\bibinfo{series}{CF '08})}.
  \bibinfo{publisher}{Association for Computing Machinery},
  \bibinfo{address}{New York, NY, USA}, \bibinfo{pages}{189–198}.
\newblock
\showISBNx{9781605580777}
\urldef\tempurl%
\url{https://doi.org/10.1145/1366230.1366265}
\showDOI{\tempurl}


\bibitem[W\"{u}rthinger et~al\mbox{.}(2012)]%
        {SelfOptimizingAst}
\bibfield{author}{\bibinfo{person}{Thomas W\"{u}rthinger},
  \bibinfo{person}{Andreas W\"{o}\ss{}}, \bibinfo{person}{Lukas Stadler},
  \bibinfo{person}{Gilles Duboscq}, \bibinfo{person}{Doug Simon}, {and}
  \bibinfo{person}{Christian Wimmer}.} \bibinfo{year}{2012}\natexlab{}.
\newblock \showarticletitle{Self-Optimizing AST Interpreters}.
\newblock \bibinfo{journal}{\emph{SIGPLAN Not.}} \bibinfo{volume}{48},
  \bibinfo{number}{2} (\bibinfo{date}{oct} \bibinfo{year}{2012}),
  \bibinfo{pages}{73–82}.
\newblock
\showISSN{0362-1340}
\urldef\tempurl%
\url{https://doi.org/10.1145/2480360.2384587}
\showDOI{\tempurl}


\bibitem[Yee et~al\mbox{.}(2010)]%
        {NativeClient}
\bibfield{author}{\bibinfo{person}{Bennet Yee}, \bibinfo{person}{David Sehr},
  \bibinfo{person}{Gregory Dardyk}, \bibinfo{person}{J.~Bradley Chen},
  \bibinfo{person}{Robert Muth}, \bibinfo{person}{Tavis Ormandy},
  \bibinfo{person}{Shiki Okasaka}, \bibinfo{person}{Neha Narula}, {and}
  \bibinfo{person}{Nicholas Fullagar}.} \bibinfo{year}{2010}\natexlab{}.
\newblock \showarticletitle{Native Client: A Sandbox for Portable, Untrusted
  X86 Native Code}.
\newblock \bibinfo{journal}{\emph{Commun. ACM}} \bibinfo{volume}{53},
  \bibinfo{number}{1} (\bibinfo{date}{jan} \bibinfo{year}{2010}),
  \bibinfo{pages}{91–99}.
\newblock
\showISSN{0001-0782}
\urldef\tempurl%
\url{https://doi.org/10.1145/1629175.1629203}
\showDOI{\tempurl}


\bibitem[Zakai(2013)]%
        {AsmJs}
\bibfield{author}{\bibinfo{person}{Alon Zakai}.}
  \bibinfo{year}{2013}\natexlab{}.
\newblock \bibinfo{title}{asm.js: an extraordinarily optimizable, low-level
  subset of JavaScript}.
\newblock \bibinfo{howpublished}{\url{http://asmjs.org}}.
\newblock
\urldef\tempurl%
\url{http://asmjs.org}
\showURL{%
\tempurl}
\newblock
\shownote{(Accessed 2021-07-29)}.


\bibitem[Zaleski et~al\mbox{.}(2005)]%
        {MixedModeContextThreading}
\bibfield{author}{\bibinfo{person}{Mathew Zaleski}, \bibinfo{person}{Marc
  Berndl}, {and} \bibinfo{person}{Angela~Demke Brown}.}
  \bibinfo{year}{2005}\natexlab{}.
\newblock \showarticletitle{Mixed Mode Execution with Context Threading}. In
  \bibinfo{booktitle}{\emph{Proceedings of the 2005 Conference of the Centre
  for Advanced Studies on Collaborative Research}} (Toranto, Ontario, Canada)
  \emph{(\bibinfo{series}{CASCON '05})}. \bibinfo{publisher}{IBM Press},
  \bibinfo{pages}{305–319}.
\newblock


\bibitem[Zhang et~al\mbox{.}(2022)]%
        {QuantifyingPythonInt}
\bibfield{author}{\bibinfo{person}{Qiang Zhang}, \bibinfo{person}{Lei Xu},
  \bibinfo{person}{Xiangyu Zhang}, {and} \bibinfo{person}{Baowen Xu}.}
  \bibinfo{year}{2022}\natexlab{}.
\newblock \showarticletitle{Quantifying the interpretation overhead of Python}.
\newblock \bibinfo{journal}{\emph{Science of Computer Programming}}
  \bibinfo{volume}{215} (\bibinfo{year}{2022}), \bibinfo{pages}{102759}.
\newblock
\showISSN{0167-6423}
\urldef\tempurl%
\url{https://doi.org/10.1016/j.scico.2021.102759}
\showDOI{\tempurl}


\bibitem[Zilli et~al\mbox{.}(2015a)]%
        {HwSwCodesign}
\bibfield{author}{\bibinfo{person}{Massimiliano Zilli},
  \bibinfo{person}{Wolfgang Raschke}, \bibinfo{person}{Reinhold Weiss},
  \bibinfo{person}{Johannes Loinig}, {and} \bibinfo{person}{Christian Steger}.}
  \bibinfo{year}{2015}\natexlab{a}.
\newblock \showarticletitle{Hardware/Software Co-Design for a High-Performance
  Java Card Interpreter in Low-End Embedded Systems}.
\newblock \bibinfo{journal}{\emph{Microprocess. Microsyst.}}
  \bibinfo{volume}{39}, \bibinfo{number}{8} (\bibinfo{date}{nov}
  \bibinfo{year}{2015}), \bibinfo{pages}{1076–1086}.
\newblock
\showISSN{0141-9331}
\urldef\tempurl%
\url{https://doi.org/10.1016/j.micpro.2015.05.004}
\showDOI{\tempurl}


\bibitem[Zilli et~al\mbox{.}(2015b)]%
        {JavaCardCompress}
\bibfield{author}{\bibinfo{person}{Massimiliano Zilli},
  \bibinfo{person}{Wolfgang Raschke}, \bibinfo{person}{Reinhold Weiss},
  \bibinfo{person}{Christian Steger}, {and} \bibinfo{person}{Johannes Loinig}.}
  \bibinfo{year}{2015}\natexlab{b}.
\newblock \showarticletitle{A Light-Weight Compression Method for Java Card
  Technology}.
\newblock \bibinfo{journal}{\emph{SIGBED Rev.}} \bibinfo{volume}{11},
  \bibinfo{number}{4} (\bibinfo{date}{jan} \bibinfo{year}{2015}),
  \bibinfo{pages}{13–18}.
\newblock
\urldef\tempurl%
\url{https://doi.org/10.1145/2724942.2724944}
\showDOI{\tempurl}


\end{thebibliography}

\appendix

\end{document}